\documentclass{ws-book9x6}
\usepackage{ws-book-thm}  
\usepackage{ws-book-har}   
\usepackage[pdfpagelabels=false,colorlinks=true,allcolors=black]{hyperref}  
\usepackage{slashed}

\title{New Phenomena and New States of Matter in the Universe}     

\makeindex

\begin{document}

\chapter[Quark deconfinement in compact stars through sexaquark condensation]{Quark deconfinement in compact stars through sexaquark condensation}
\label{ch1}

\author[David Blaschke,
Oleksii Ivanytskyi, and
Mahboubeh Shahrbaf]{David Blaschke$^*$, 
Oleksii Ivanytskyi$^\dagger$ and  
Mahboubeh Shahrbaf$^\ddagger$
}
\address{Institute of Theoretical Physics, 
	University of Wroclaw 	\\
	pl. M. Born 9, 50-204 
	Wroclaw, Poland\\
    $^*$david.blaschke@uwr.edu.pl\\   
    $^\dagger$oleksii.ivanytskyi@uwr.edu.pl\\
    $^\ddagger$mahboubeh.shahrbafmotlagh@uwr.edu.pl}

\section*{Abstract}
\label{abstract}
In this contribution, we present for the first time a scenario according to which early quark deconfinement in compact stars is triggered by the Bose-Einstein condensation (BEC) of a light sexaquark (S) with a mass $m_S<2054$ MeV, that has been suggested as a  candidate particle to explain the baryonic dark matter in the Universe. 
The onset of S BEC marks the maximum mass of hadronic neutron stars and it occurs when the condition for the baryon chemical potential $\mu_b=m_S/2$ is fulfilled in the center of the star, corresponding to $M_{\rm onset}\lesssim 0.7~M_\odot$. 
In the gravitational field of the star the density of the BEC of 
the S increases until a new state of the matter is attained, 
where each of the S-states got dissociated into a triplet of color-flavor-locked (CFL) diquark states. 
These diquarks are the Cooper pairs in the color superconducting CFL phase of quark matter, so that the developed scenario corresponds to a 
BEC-BCS transition in strongly interacting matter.
For the description of the CFL phase, we develop here for the first time the three-flavor extension of the density-functional formulation of a chirally symmetric Lagrangian model of quark matter where confining properties are encoded in a divergence of the scalar self-energy at low densities and temperatures. 

\section{Introduction}
\label{sec1.1}

The discussion of quark matter in compact stars has a long history. 
It started with the early works by 
\citet{Ivanenko:1965dg} and 
\citet{Itoh:1970uw} 
but a proper foundation for the necessity of quark deconfinement in QCD existed only after asymptotic freedom was proven. Based on this argument, 
\citet{Baym:1976yu}
developed the concept of a thermodynamic bag model in order to describe macroscopic volumes of deconfined quark matter, for instance inside neutron stars. 
At that time also the idea of a third family of compact stars was already formulated by 
\citet{Gerlach:1968zz} but found practically no resonance at that time.

Another idea was that of "collapsed nuclei" \citep{Bodmer:1971we}, namely that much smaller and deeper bound nuclei, possibly electrically neutral due to their hypercharge, could exist in nature, separated from ordinary nuclei by a sufficiently high barrier to guarantee stability of the latter on cosmological time scales.

After the MIT bag model was conceived \cite{Chodos:1974je} and used to explain the structure of baryons \cite{Chodos:1974pn}, a hypothetical hypernucleus with baryon number $6$ and strangeness $-6$ (hexalambda) was proposed by 
\citet{Terazawa:1979} within the MIT bag model.

The idea of an absolutely stable strange quark matter state with astrophysical applications was formulated by Witten as "Cosmic Separation of Phases" \citep{Witten:1984rs} and immediately developed further by 
\citet{Farhi:1984qu} as "Strange Matter". 
Despite intense searches for remnants of strange matter at different length scales, in the Cosmos and in Laboratory experiments, no conclusive evidence for the existence of strange quark matter nuggets has been found yet.

About a decade ago it was even argued that strangeness is likely absent from neutron star interiors \cite{Blaschke:2010vd}. 
The reasoning for this was based on 
the fact that a standard solution for the hyperon puzzle was early quark deconfinement \cite{Baldo:2003vx,Burgio:2002sn}
(for a recent update fulfilling the $2~M_\odot$ constraint, see \cite{Shahrbaf:2019vtf,Shahrbaf:2020uau}) which eliminated the appearance of hyperons as scenario for compact star interiors. 
When for the quark matter phase a Nambu--Jona-Lasinio (NJL) type model with its sequential occurrence of quark flavors \cite{Blaschke:2008br} was adopted, then after the possible one- and two-flavor color superconducting phases, the occurrence of the strange quark flavor in the CFL color superconducting phase was accompanied with the gravitational instability of the corresponding hybrid star configurations, see also
\cite{Klahn:2006iw,Klahn:2013kga}.
With this line of reasoning and the demonstration of numerous examples, there was no place for strangeness in compact stars, neither in the hadronic nor in the quark matter phase.

Concerning the idea of absolutely stable strange quark matter
\cite{Bodmer:1971we,Witten:1984rs,Farhi:1984qu} and its realization in the form of strange stars \cite{Alcock:1986hz,Haensel:1986qb}, one could even formulate a "No-Go" conjecture for the absolute stability of strange quark matter, at least for a description within NJL-type models \cite{Klahn:2017cyo}.

In our contribution to this book we want to reopen the chapter 
of strange quark matter in compact stars, based on the new, multi-messenger phenomenology of compact stars and the possibility of a light sexaquark as a dark matter particle that evaded detection in laboratory experiments due to its stability against decays on cosmological timescales. 
This new perspective is made possible by a recent development of a density functional approach to quark matter in compact stars which allows to address confining effects \cite{Kaltenborn:2017hus} that were absent in the formulation using NJL models of quark matter.
In \cite{Ivanytskyi:2021dgq,Ivanytskyi:2022oxv}, this approach was refined so that its Lagrangian was manifestly chirally symmetric and the important diquark interactions resulting in color superconducting phases were added.
In the present contribution, for the first time, this approach will be generalized to three quark flavors and we will show that the light sexaquark acts as a trigger for entering the CFL quark matter phase at an onset mass below $0.7~M_\odot$, while fulfilling the modern mass-radius constraints on pulsars.


\section{Bose-Einstein condensation of sexaquarks as a trigger for strange quark matter in compact stars}

The discussion of a light and compact sexaquark (S) state with the quark content (uuddss)
and the consequences for neutron star phenomenology that follow from its possible existence is of much interest in nuclear physics, particle physics as well as astrophysics. 
The S is an electrically neutral spin-less boson with baryon number $B_S=2$ and strangeness $S_S=-2$ in a flavor-singlet state.
For $m_S \leq 2 (m_p + m_e) = 1877.966$~MeV, due to baryon number conservation the S is stable, while for 
$m_S \leq m_p + m_e + m_\Lambda = 2054.466$~MeV, it should decay with a lifetime exceeding the age of the universe \cite{Farrar:2003qy}. 

In a recent work by Shahrbaf et al. 
\cite{Shahrbaf:2022upc} it was shown that the existence of a light S in the mass range $1885~\mbox{MeV} < m_S < 2054$~MeV leads to S BEC in neutron stars.
For the upper bound of this mass range, i.e. $m_S=2054$ MeV, the condensation happens at masses as low as $M\approx 0.7~M_\odot$ limiting thus the maximum mass of neutron stars.
It was also shown that a plausible positive mass shift with increasing density allows for a stable sequence of neutron stars with sexaquark enriched matter in their cores that eventually undergoes a deconfinement transition to color superconducting (two-flavor) quark matter.

Such hybrid stars with cores of deconfined, color superconducting quark matter would then populate the mass range from the typical neutron star masses $\sim 1.4~M_\odot$ up to the maximum mass $M_{\rm max}>2~M_\odot$. 
The precise value of the maximum mass and the radius where it is reached depend on the details of the choice of parameters for the quark matter model.

Here we will examine a new scenario where the S does not undergo a mass shift and therefore its BEC at a low star mass is inevitable and leads to a collapse of the hadronic neutron star which gets stopped by quark deconfinement and the formation of a hybrid star.
This new, stable hybrid star sequence will fulfill the recent constraints on neutron star radii at $1.4~M_\odot$ from the gravitational wave signal of the binary neutron star merger GW170817 by the LIGO-Virgo Collaboration \cite{LIGOScientific:2018cki} and at $2.0~M_\odot$ by the recent NICER measurement on PSR J0740+6620 \cite{Miller:2021qha,Riley:2021pdl}. 

This new scenario is built on the existence of the S as a deeply bound state with low enough mass to be stable on cosmological time scales (age of the Universe) and is therefore a dark matter (DM) candidate. The observed DM to baryon ratio is $\Omega_{DM}/\Omega_B = 5.3\pm 0.1$ \cite{Planck:2015fie, ParticleDataGroup:2018ovx} and a successful model for DM has to reproduce this value. 
An abundance of S DM (SDM) in agreement with this observation has been obtained within a statistical model on the basis of assumptions for the quark masses and an effective temperature $T_{\rm eff}=156$ MeV \footnote{This value is motivated by the recent result of $T_c=156.5 \pm 1.5$ MeV for the pseudocritical temperature obtained in lattice QCD simulations \cite{HotQCD:2018pds}} of the transition from the quark-gluon plasma to the hadronic phase when $m_S=1860$ MeV
\cite{Farrar:2020zeo}. 
When $m_S=2m_p=1876.54$ MeV, the observed ratio $\Omega_{SDM}/\Omega_B = 5.3$ is obtained for $T_{\rm eff}=150$ MeV.
It was first proposed in \cite{Farrar:2002ic} that S is a candidate of DM and this idea was followed during the next years as well \cite{Farrar:2017eqq, Farrar:2017ysn, Farrar:2018hac,Farrar:2020zeo}.


The fact that the light S cannot decay and that it is electrically neutral explains why it has so far evaded detection in laboratory experiments \cite{Farrar:2020zeo}. 

In the present contribution, the mass of the S is considered to be constant and set equal to the upper bound of its stability range, $m_S=2054$ MeV. At this mass, the S is light enough to be
metastable on cosmological scales so as to serve as a DM candidate, and sufficiently heavy on the other hand to not disturb the stability of nuclei \cite{Farrar:2003gh,Farrar:2003qy,Gross:2018ivp}.

We want to consider the possible relevance of S for the properties of neutron stars based on the recent multi-messenger observations that constrain their regions of accessibility in the mass-radius diagram. When the S is found in cold dense baryonic matter in neutron stars, it may undergo BEC as soon as the baryochemical potential in the center of the star fulfills $\mu_B=m_S/2$. 
For $m_S=2054$ MeV this occurs for a star with $M=0.7~M_\odot$ and, due to a saturation of the pressure with the BEC, this value marks the maximum mass that can be reached. 
This would be in clear contradiction with the observation of pulsars as massive as $2~M_\odot$  like PSR J0740+6620 
\cite{Fonseca:2021wxt} or PSR J0348+0432 \cite{Antoniadis:2013pzd}. 
Therefore, we introduce the sexaquark dilemma in this section and its solution by quark deconfinement.

The EoS of hadronic matter is obtained from a generalized relativistic density functional (GRDF) with baryon-meson couplings that depend on the total baryon density of the system. The original density functional for nucleonic matter considers the isoscalar $\sigma$ and $\omega$ mesons and the isovector $\rho$ meson as exchange particles that describe the effective in-medium interaction. The density dependence of the couplings is adjusted to describe properties of atomic nuclei \cite{Typel:1999yq,Typel:2005ba}. It has been confirmed that such GRDFs are successful in reproducing the properties of nuclear matter around nuclear saturation \cite{Klahn:2006ir}.
GRDFs with different parameterization of the density dependent couplings have been studied in \cite{Typel:2018cap}.
In the present work the parameterization DD2 \cite{Typel:2009sy} is used for the $\sigma$, $\omega$, and $\rho$ couplings. It predicts characteristic nuclear matter parameters that are consistent with recent constraints \cite{Oertel:2016bki} and leads to a rather stiff EoS at high baryon densities with a maximum neutron star mass of $2.4~M_\odot$  in a pure nucleonic scenario of the strongly interacting system without hyperons.

In the description of neutron-star matter one has to consider that new baryonic degrees of freedom can become active with increasing density as the chemical potentials rise. Neglecting any interaction, a new species appears when the corresponding chemical potential crosses the particle mass.

With the GDRF for hadronic matter at hand it is possible to explore the EoS and corresponding properties of neutron stars. There are different scenarios to be distinguished in the following. The most simple case with only nucleons and leptons corresponds to the original DD2 model as presented in \cite{Typel:2009sy}. Adding hyperons the model is called DD2Y-T  to distinguish it from the similar model DD2Y introduced in \cite{Marques:2017zju}. The DD2Y-T predictions were compared already to other EoS models with hyperons in \cite{Stone:2019blq}.
 
Finally, after including also the sexaquark, there is the full model which will be denoted DD2Y-T+S$_{2054}$ in the following. 
When the mass of the S is $m_S=2054$ MeV, the onset of the S occurs at $n_b^{\rm onset, S}=0.25$ fm$^{-3}$ (corresponding to more than $1.5$ times the saturation density) with an immediate appearance of the BEC and a collapse of the neutron star because with increasing central density the pressure remains constant.

To show that the sexaquark onset triggers a first-order phase transition to a color superconducting quark matter, we need the EoS for quark matter.
There is a phenomenological  formulation of the EoS of quark matter in use which has been introduced and motivated in Ref.~\cite{Alford:2004pf}. 
In that work, the EoS of symmetric quark matter consists of the first three terms of a series in even powers of the quark chemical potential
\begin{equation}
\label{eq:quartic}
\Omega_{QM}=-\frac{3}{4\pi^2} a_4 \mu^4 + \frac{3}{4\pi^2} a_2 \mu^2 + B_{\textrm{eff}}~,
\end{equation}
where $a_4$, $a_2$, and $B_{\textrm{eff}}$ are coefficients independent of $\mu$.

The quartic coefficient $a_4=1-c$ is well defined for an ideal massless gas for which c=0. 
Perturbative QCD corrections in lowest order, i.e. $O(\alpha_s)$ for massless quark matter lead to a reduction of $a_4$, e.g., accounted for by $c=0.3$ so that $a_4=0.7$.
The quadratic $\mu^2$ term arises from an expansion in the finite strange quark mass $m_s$ and the diquark pairing gap
$\Delta$, so that $a_2 = m_s^2-4\Delta^2$. 
In CFL quark matter, they are almost in the same order of about $100$ MeV so that the coefficient $a_2$ is almost zero and this corresponds to a constant speed of sound $c_s^2=1/3$, the conformal limit \cite{Alford:2004pf}.
For simplicity we set $m_s=0$, providing electric neutrality of the considered quark matter EoS without leptons.
The values of coefficients in \eqref{eq:quartic} which are used in this work are taken from \cite{Antic:2021zbn}
and amount to $a_4 = 0.22 $, $a_2 =-(299.6 ~{\rm MeV})^2$, $ B_{\textrm{eff}}^{1/4} = 174.2$ MeV.

\begin{figure}[!h]
\centering
	\includegraphics[width=0.7\textwidth]{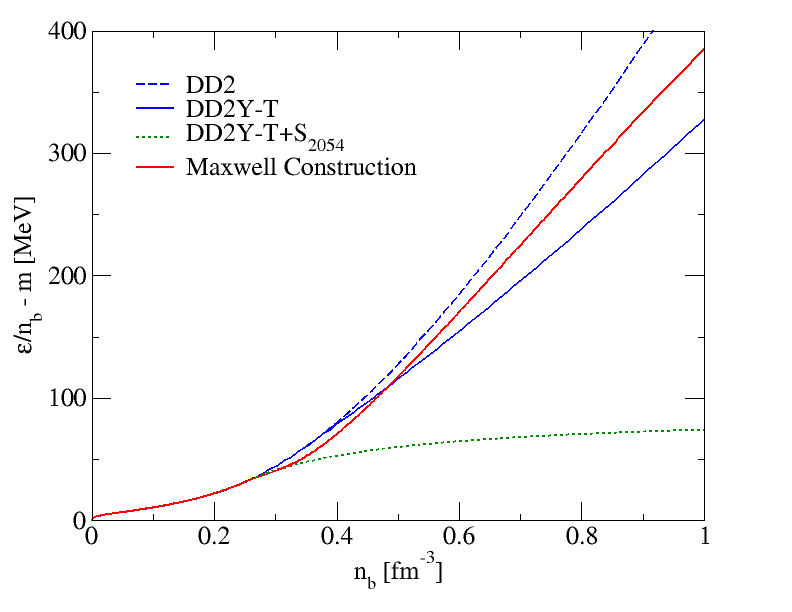}
	\caption{Energy per particle $\varepsilon/n_b -m$ as a function of baryon density $n_b$ for 
	DD2, DD2Y-T and DD2Y-T+S$_{2054}$. 
	The red solid line shows the hybrid EoS resulting from a Mexwell construction between DD2 and the used quark matter EoS. 
		\label{fig:EperA}
	}
\end{figure}

In the present work, devoted to the application to neutron stars, we can restrict ourselves to the case of matter at zero temperature. 
At densities below saturation there are no hyperons or sexaquarks and the unified crust EoS of the original GRDF-DD2 model with clusters is used. It contains the well-known sequence of nuclei in a body-centered cubic lattice with a uniform background of electrons and a neutron gas above the neutron drip line. 
The transition to homogeneous matter just below the nuclear saturation density is described consistently within the same approach. The main modification is to include the new degrees of freedom at supersaturation densities in the GRDF model.
We show the energy per particle  $\varepsilon/n_b -m$, where $m$ is the nucleon mass, as a function of baryon density $n_b$ for different scenarios in hadronic matter as well as a used quark matter EoS in CFL phase in Fig.~\ref{fig:EperA}.

\begin{figure}[!h]
	\includegraphics[width=0.49\textwidth]{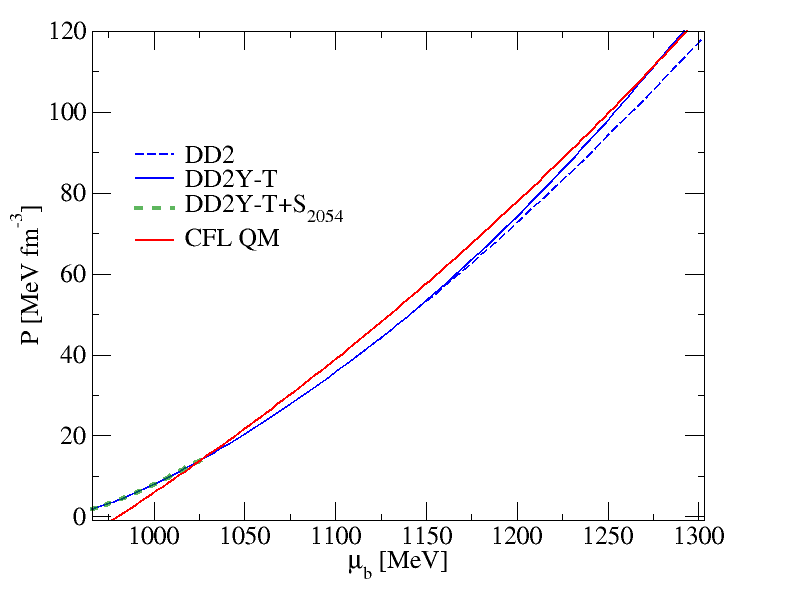}
	\includegraphics[width=0.49\textwidth]{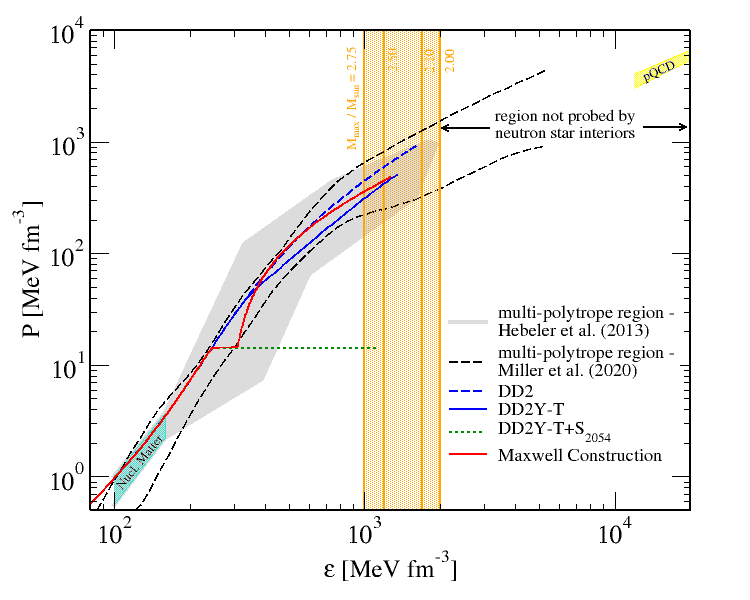}
	\caption{Pressure $P$ for DD2, DD2Y-T and DD2Y-T+S$_{2054}$ as a function of the baryochemical potential $\mu_b$ (left panel)
	and as a function of energy density $\varepsilon = \mu_b n_b - P$ (right panel), where we 
	emphasize that the region between highest energy densities in neutron stars and applicability of perturbative QCD can not be probed with neutron stars.
	Line styles as in Fig.~\ref{fig:EperA}.
	The red solid line corresponds to a hybrid EoS for which the sexaquark onset triggers a first-order phase transition to a color superconducting quark matter that is fitted to CFL phase \cite{Shahrbaf:2022upc}.
	We show as grey hatched region the EoS constraint from \cite{Hebeler:2013nza} and by black dashed lines the one from \cite{Miller:2019nzo}.	
		\label{fig:eos1a1b}
	}
\end{figure}

\begin{figure}[!h]
	\includegraphics[width=0.47\textwidth]{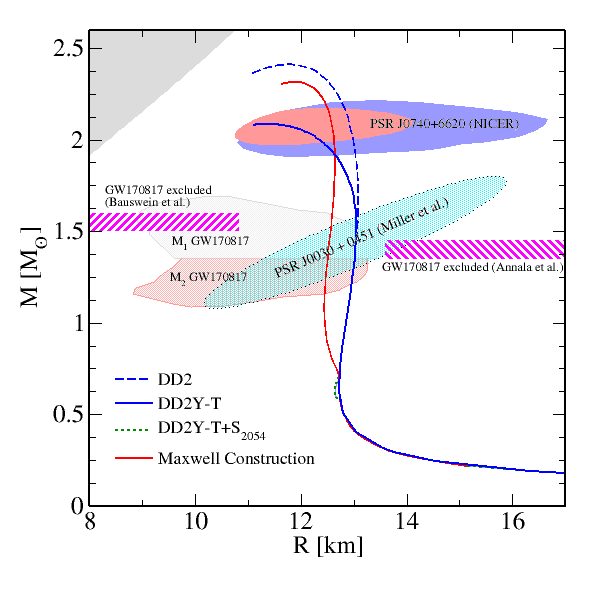}
	\includegraphics[width=0.47\textwidth]{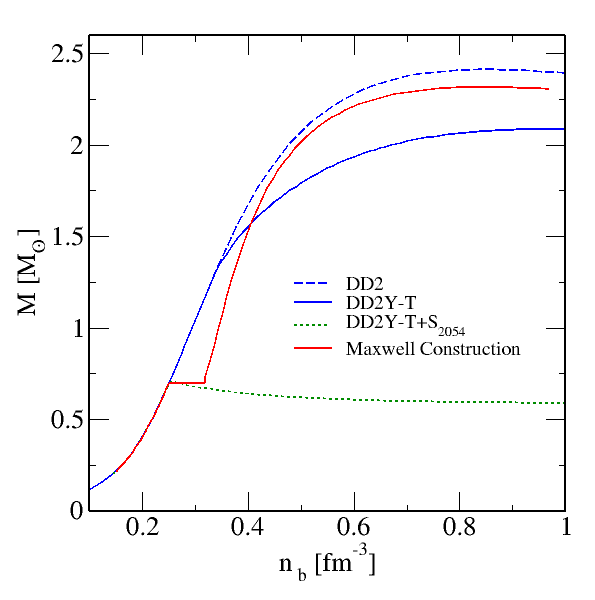}
	\caption{Mass vs radius (left panel) and mass vs. baryon density (right panel) for compact stars in which the sexaquark S particle is assumed with a constant mass 
	$m_S=2054$ MeV (solid orange line). 
	The black solid line corresponds to the hadronic EoS model without sexaquark while the black dashed line stands for the hadronic model without hyperons. 
	The red solid line corresponds to a hybrid EoS for which the sexaquark onset triggers a first-order phase transition to a color superconducting quark matter that is fitted to CFL phase EoS \cite{Shahrbaf:2022upc}.
	For a comparison the new $1.0~\sigma$ mass-radius constraints from the NICER analysis of observations of the massive pulsar PSR J0740+6620 \cite{Fonseca:2021wxt} are indicated in red \cite{Riley:2021pdl} and blue \cite{Miller:2021qha} regions. Additionally, the magenta bars mark the excluded regions for a lower limit \cite{Bauswein:2017vtn} and an upper limit \cite{Annala:2017llu} on the radius deduced from the gravitational wave observation GW170817. 
	The light bluee region is from the NICER mass-radius measurement on PSR J0030+0451 \cite{Miller:2019cac}.
		\label{fig:Mx} 
	}
\end{figure}
The pressure $P = - \Omega_{QM}$ as a function of baryonic chemical potential $\mu_b=3\mu$ and energy density for DD2, DD2Y-T, DD2Y-T+S$_{2054}$ as well as the deconfined quark matter are shown in Fig. \ref{fig:eos1a1b}. 
The corresponding mass-radius and mass-density curves are shown in Fig. \ref{fig:Mx}.
We recognize the instability of the hadronic model that includes the S by the fact that the square of the eigenmode of radial oscillations of the spherical star, given by $\omega^2=dM/dn_b^c$,
is negative for $n_b^c>n_b^{\rm onset, S}$.
Here $M$ denotes the mass of the star, $n_b^c$ is the baryon density at its center and $n_b^{\rm onset, S}$
is the baryon density at the sexaquark onset.
The red solid line in Figs. \ref{fig:eos1a1b} and \ref{fig:Mx} corresponds to a hybrid EoS for which the sexaquark onset triggers a first-order phase transition to a color superconducting quark matter phase that is fitted by a CSS form of EoS \cite{Shahrbaf:2022upc}.
In the next section, we will consider a new density functional approach to color superconducting strange quark matter which is capable of addressing quark confinement by a divergent scalar selfenergy (quark mass) following from the suggested chirally symmetric energy density functional model. 

\section{Density functional approach to strange quark matter deconfinement}
\label{sec1.3}

Here we present the 3-flavor extension of the  density functional approach to confining quark matter that was initiated by \cite{Kaltenborn:2017hus}
and then generalized to a chirally-symmetric 
Lagrangian formulation in \cite{Ivanytskyi:2021dgq,Ivanytskyi:2022oxv}, where also the effect of color superconductivity was included. The confining aspect of quark matter is introduced to the consideration via a fast growth of the quark self-energy already at the mean-field level. This simple mechanism provides an efficient suppression of quark degrees of freedom in the confined region of phase diagram. In principle, the approach allows even divergence of the quark-self energy, which corresponds to an absolute ``confinement''. However, the analysis performed in Ref. \cite{Ivanytskyi:2021dgq} as well as in \cite{Ivanytskyi:2022oxv} suggests quite high but finite values of the mean-field self-energy of confined quarks.

The mathematical formulation of the density functional approach to three-flavor three-color quark matter is based on the Lagrangian
\begin{eqnarray}
\label{I_O}
\mathcal{L}=\overline{q}(i\slashed{\partial}-m)q
+\mathcal{L}_V+\mathcal{L}_D-
\mathcal{U},
\end{eqnarray}
written for three-component flavor spinor of quarks $q^T=(u,d,s)$. The diagonal matrix $m={\rm diag}(m_u,m_d,m_s)$ acts in the flavor space and represents current quark masses. The second term in Eq. (\ref{I_O}) generates the vector repulsion 
\begin{eqnarray}
\label{II_O}
\mathcal{L}_V=-G_V(\overline{q}\gamma_\mu q)^2.
\end{eqnarray}
Its strength is controlled by the coupling constant $G_V$. 
This repulsive interaction can be motivated by the non-perturbative gluon exchange \cite{PhysRevD.100.034018} and is phenomenologically important in order to provide enough stiffness of dense quark matter needed to support existence of hybrid quark-hadron stars of two solar masses. The third term in Eq. (\ref{I_O}) stands for scalar color-flavor antitriplet attractive interaction among quarks
\begin{eqnarray}
\label{III_O}
\mathcal{L}_D&=&G_D\sum_{a,b=2,5,7}
(\overline{q}i\gamma_5\mathcal{T}_a\lambda_b q^c)(\overline{q}^c i\gamma_5\mathcal{T}_a\lambda_bq),
\end{eqnarray}
where $G_D$ is coupling constant of the color antitriplet scalar diquark channel, where the charge conjugate quark field is defined as $q^c=i\gamma_2\gamma_0\overline{q}^T$, while eight flavor $\mathcal{T}_a$ and eight color $\lambda_b$ Gell-Mann matrices are connected to the generators of the corresponding SU(3) groups. 
For further convenience we also introduce $\mathcal{T}_0=\sqrt{\frac{2}{3}}$. 
We note that the summation in Eq. (\ref{III_O}) is performed only over the antisymmetric generators with indices $a,b=2,5,7$. 
The quark interaction in this channel is responsible for the diquark pairing and formation of the color superconducting phases of quark matter. 

The energy density functional $\mathcal{U}$ models the effects related to dynamical restoration or breaking of chiral symmetry of quark matter. Following the symmetries of the QCD Lagrangian we require $\mathcal{U}$ to be chirally symmetric. The most obvious way to fulfill this requirement corresponds to choosing the argument of $\mathcal{U}$ to be chirally symmetric itself. The first part of such argument is 
\begin{eqnarray}
\label{IV_O}
\mathcal{O}_1=\frac{1}{2}\sum_{a=0}^8\Bigl((\overline{q}\mathcal{T}_aq)^2+
(\overline{q}i\gamma_5\mathcal{T}_aq)^2\Bigl).
\end{eqnarray}
We also introduce the instanton induced term of the 't Hooft form
\begin{eqnarray}
\label{V_O}
\mathcal{O}_2=\zeta\Bigl(\det\bigl(\overline{q}(1+\gamma_5)q\bigl)+
\det\bigl(\overline{q}(1-\gamma_5)q\bigl)\Bigl),
\end{eqnarray}
with $\zeta$ being a constant discussed below and determinant carried in the flavor space. Thus, similar to the two-flavor case \cite{Ivanytskyi:2021dgq,Ivanytskyi:2022oxv} we define the chirally symmetric density functional as
\begin{eqnarray}
\label{VI_O}
\mathcal{U}=D_0\left[X-\mathcal{O}_1-\mathcal{O}_2\right]^{\frac{1}{3}},
\end{eqnarray}
where $D_0$ and $X$ are constants. The form of this potential is motivated by the string flip model \cite{Horowitz:1985tx,Ropke:1986qs}.

The present consideration is limited to the mean-field case, when expectation values of the operators are 
\begin{eqnarray}
\label{VII_O}
\langle\overline{f}f'\rangle&=&\langle\overline{f}f\rangle\delta_{ff'},\\
\label{VIII_O}
\langle\overline{f}i\gamma_5\mathcal{T}_af'\rangle&=&0.
\end{eqnarray}
Hereafter $f=u,d,s$ is the quark flavor index.
Within this approximation flavor matrices $\overline{q}(1\pm\gamma_5)q$ become diagonal, i.e.
\begin{eqnarray}
\label{IX_O}
\langle\overline{q}(1\pm\gamma_5)q\rangle&=&{\rm diag}
(\langle\overline{u}u\rangle,\langle\overline{d}d\rangle,
\langle\overline{s}s\rangle).
\end{eqnarray}
Using Eqs. (\ref{VII_O}) - (\ref{IX_O}) we find the mean-field interaction potential (\ref{VI_O})
\begin{eqnarray}
\label{X_O}
\mathcal{U}^{(0)}=D_0\left[X-\sum_f\langle\overline{f}f\rangle^2-2\zeta\prod_f\langle\overline{f}f\rangle\right]^{\frac{1}{3}}.
\end{eqnarray}
Let us analyze this expression in the two-flavor case. For this the strange quark mass should be approached to infinity $m_s\rightarrow\infty$ leading to $\langle\overline{s}s\rangle=\langle\overline{s}s\rangle_0$ and
\begin{eqnarray}
\label{XI_O}
\mathcal{U}^{(0)}_{m_s\rightarrow\infty}=D_0\Biggl[X-\langle\overline{s}s\rangle_0^2-
\Biggl(\langle\overline{u}u\rangle^2+\langle\overline{d}d\rangle^2+2\langle\overline{u}u\rangle\langle\overline{d}d\rangle\cdot\zeta\langle\overline{s}s\rangle_0\Biggl)\Biggl]^{\frac{1}{3}}.\quad~~
\end{eqnarray}
Hereafter the subscript index ``$0$'' denotes the quantities defined in the vacuum. 
Obviously, the potential given by Eq. (\ref{XI_O}) should coincide with the two-flavor one from Ref. \cite{Ivanytskyi:2021dgq,Ivanytskyi:2022oxv}
\begin{eqnarray}
\label{XII_O}
\mathcal{U}^{(0)}_{N_f=2}=D_0\Biggl[(1+\alpha)
\Bigl(\langle\overline{u}u\rangle_0+\langle\overline{d}d\rangle_0\Bigl)^2-
\Bigl(\langle\overline{u}u\rangle+\langle\overline{d}d\rangle\Bigl)^2\Biggl]^{\frac{1}{3}}
\end{eqnarray}
with $\alpha=const$. From this we immediately express the parameters of the tree-flavor potential through the ones of the two-flavor potential
\begin{eqnarray}
\label{XIII_O}
X&=&(1+\alpha)
\Bigl(\langle\overline{u}u\rangle_0+\langle\overline{d}d\rangle_0\Bigl)^2+\langle\overline{s}s\rangle_0^2,\\
\label{XIV_O}
\zeta&=&\frac{1}{\langle\overline{s}s\rangle_0}.
\end{eqnarray}

Having the chirally symmetric interaction potential defined we can analyze the modification of the single quark properties caused by the interaction described by $\mathcal{U}$. 
For this we follow the strategy of Ref. \cite{Kaltenborn:2017hus} and perform the first order expansion around the mean-field expectation values of scalar operators $\overline{f}f$. 
We point out that the expansion around the expectation values of the pseudoscalar operators $\overline{q}i\gamma_5\mathcal{T}_aq$ produces non-vanishing terms starting from the second order only \cite{Ivanytskyi:2021dgq,Ivanytskyi:2022oxv}. 
The first order terms include derivatives of the mean-field potential with respect to chiral condensates of different flavors, which are nothing else as the flavor matrix of the mean-field self-energy of quarks 
\begin{eqnarray}
\label{XV_O}\Sigma^{(0)}&=&{\rm diag}(\Sigma_u^{(0)},\Sigma_d^{(0)},\Sigma_s^{(0)}),\quad\Sigma_f^{(0)}\equiv\frac{\partial\mathcal{U}^{(0)}}{\partial \langle\overline{f}f\rangle}.
\end{eqnarray}
With this notation the first-order expanded interaction potential and the corresponding Lagrangian become
\begin{eqnarray}
\label{XVI_O}
\mathcal{U}^{(1)}&=&\mathcal{U}^{(0)}+\overline{q}\Sigma^{(0)} q-\langle\overline{q}\Sigma^{(0)}  q\rangle,\\
\label{XVII_O}
\mathcal{L}^{(1)}&=&\overline{q}(i\slashed{\partial}- m^*)q+\mathcal{L}_V+\mathcal{L}_D-\mathcal{U}^{(0)}+\langle\overline{q}\Sigma^{(0)}  q\rangle.
\end{eqnarray}
The latter includes the matrix of medium-dependent quark masses
\begin{eqnarray}
\label{XVIII_O}
m^*=m+\Sigma^{(0)} .
\end{eqnarray}
In the region of small temperatures and densities, the mean-field value of the argument of the interaction potential (\ref{VI_O}) can be approximated as $X-\mathcal{O}_1-\mathcal{O}_2\sim\langle q^+ q\rangle.$

It is interesting to consider the effective quark mass in the vacuum
\begin{eqnarray}
\label{XIX_O}
m_f^*=m_f-\frac{2D_0\alpha^{-\frac{2}{3}}}{3\left[\langle\overline{u}u\rangle_0+\langle\overline{d}d\rangle_0
\right]^{\frac{4}{3}}}
\left[\langle\overline{f}f\rangle_0+\frac{1}{\langle\overline{s}s\rangle_0}\prod\limits_{f'\neq f}\langle\overline{f'}f'\rangle_0\right].
\end{eqnarray}
From this expression it becomes clear that a vanishing $\alpha$ leads to a divergent effective mass of quarks at $\langle\overline{f}f\rangle=\langle\overline{f}f\rangle_0$ and, consequently, to an absolute suppression of quarks in the confining region. 
A small but finite value of $\alpha$ leads to a large $m_f^*$, which is sufficient for an efficient suppression of quarks at low temperatures and densities. 
Below, however, we consider the limiting case of $\alpha=0$, which allows us to demonstrate the confining mechanism of the present approach in the most radical way.

Mean-field treatment of the vector and diquark pairing channels corresponds to the linearization of the Lagrangian around the expectation values of operators $\overline{q}\gamma_\mu q$ and $\overline{q}^c i\gamma_5\mathcal{T}_a\lambda_bq$. 
Only the $\mu=0$ component of the first of them yields a non-vanishing expectation value $\langle q^+ q\rangle$ being the quark number density. 
The diquark pairing operators generate three non-vanishing condensates  
\begin{eqnarray}
\label{XX_O}
\langle \overline{q}^c i\gamma_5\mathcal{T}_2 \lambda_2q\rangle,\quad
\langle \overline{q}^c i\gamma_5\mathcal{T}_5 \lambda_5q\rangle,\quad
\langle \overline{q}^c i\gamma_5\mathcal{T}_7\lambda_7q\rangle.
\end{eqnarray}
They correspond to the formation of the following diquark pairs: $(u_r,d_g)$ with $(u_g,d_r)$, $(d_g,s_b)$ with $(d_b,s_g)$, and $(s_b,u_r$ with $(s_r,u_b)$, respectively. Here the subscript index running over $(r,g,b)$ represents the color state of quarks.
The resulting mean-field Lagrangian is
\begin{eqnarray}
\mathcal{L}_{MF}+q^+\mu q&=&{\overline Q}\mathcal{S}^{-1} Q\nonumber\\
\label{XXI_O}
&+&G_V\langle q^+ q\rangle^2-\sum_{a=2,5,7}|\langle \overline{q}^c i\gamma_5\mathcal{T}_a\lambda_aq\rangle|^2-\mathcal{U}^{(0)}+\langle\overline{q}\Sigma^{(0)} q\rangle.\quad
\end{eqnarray}
It is written through the Nambu-Gorkov bispinor $Q=\frac{1}{\sqrt{2}}\left(\begin{array}{l}q\\q^c\end{array}\right)$ and the propagator
\begin{eqnarray}
\label{XXII_O}
\mathcal{S}^{-1}=\left(
\begin{array}{l}
\hspace*{1cm}i\slashed{\partial}-m^*+\mu^*\gamma_0\hspace*{1cm}
\sum\limits_{a=2,5,7}i\gamma_5\mathcal{T}_a\lambda_a\langle \overline{q}^c i\gamma_5\mathcal{T}_a\lambda_aq\rangle\\
\hspace*{-.2cm}\sum\limits_{a=2,5,7}i\gamma_5\mathcal{T}_a\lambda_a\langle \overline{q}^c i\gamma_5\mathcal{T}_a\lambda_aq\rangle^*\hspace*{1.3cm}
i\slashed{\partial}-m^*-\mu^*\gamma_0
\end{array}\right).\quad
\end{eqnarray}
We note that the vector repulsion renormalizes the quark chemical potentials as
\begin{eqnarray}
\label{XXIII_O}
\mu^*&=&{\rm diag}(\mu_u,\mu_d,\mu_c)-2G_V\langle q^+q\rangle.
\end{eqnarray}
The Lagrangian (\ref{XXI_O}) is quadratic in the quark fields allowing their functional integration and leading to the thermodynamic potential
\begin{eqnarray}
\label{XXIV_O}
\Omega&=&-\frac{T}{2V}{\rm Tr}\ln(\beta\mathcal{S}^{-1})
\nonumber\\
&&-G_V\langle q^+ q\rangle^2+G_D\sum_{a=2,5,7}|\langle \overline{q}^c i\gamma_5\mathcal{T}_a\lambda_a q\rangle|^2+\mathcal{U}^{(0)}-\langle\overline{q}\Sigma^{(0)} q\rangle,
\end{eqnarray}
where $\beta=\frac{1}{T}$ is the inverse temperature, $V$ stands for the system volume and the trace is performed over the Nambu-Gorkov, Dirac, color, flavor, momentum and Matsubara indices. The latter appear since quark propagator in Eq. (\ref{XXIV_O}) is given in the momentum representation. 
Solving this trace requires eigenvalues of $\mathcal{S}^{-1}$, which in the Nambu-Gorkov-Dirac-color-flavor space is a $72\times 72$ matrix. 
In the case of equal current quark masses and chemical potentials these eigenvalues are highly degenerate provided the fact that the effective quark masses, the chiral condensates of three quark flavors and the three non-vanishing diquark condensates coincide. 
For simplicity, we consider below such a regime of the present model. 
In order to make the consideration as introductory as possible we also neglect current quark masses at all. 
This leads to the effective quark mass
\begin{eqnarray}
\label{XXV_O}
m^*=\sigma\cdot\left(\frac{D_0}{\sigma_0^2}\right)^{\frac{2}{3}}\cdot\frac{2\left[1+\frac{\sigma}{\sigma_0}\right]}{3\left[5-\frac{3\sigma^2}{{\sigma_0}^2}-\frac{2\sigma^3}{\sigma_0^3}
\right]^{\frac{2}{3}}}
\end{eqnarray}
expressed through the reduced mass parameter 
\begin{eqnarray}
\label{XXVI_O}
\sigma\equiv-\frac{\langle \overline{u}u\rangle}{D_0}=-\frac{\langle \overline{d}d\rangle}{D_0}=-\frac{\langle \overline{s}s\rangle}{D_0}.
\end{eqnarray}
In the color-flavor space the eigenvalues of the inverse Nambu-Gorkov propagator split into an octet and a singlet. 
They are
\begin{eqnarray}
\label{XXVII_O}
\omega_{\rm oct}^\pm&=&\sqrt{(\omega^\pm)^2+\Delta^2},\\
\label{XXVIII_O}
\omega_{\rm sing}^\pm&=&\sqrt{(\omega^\pm)^2+(2\Delta)^2},
\end{eqnarray}
where $\omega^\pm=\omega\pm\mu^*$, 
$\omega=\sqrt{k^2+m^{*2}}$, $k$ is the quark momentum, $\mu^*=\mu_u^*=\mu_d^*=\mu_s^*$, and the diquark pairing gap is defined as
\begin{eqnarray}
\label{XXIX_O}
\Delta\equiv2G_D|\langle \overline{q}^c i\gamma_5\mathcal{T}_2 \lambda_2q\rangle|=2G_D|\langle \overline{q}^c i\gamma_5\mathcal{T}_5 \lambda_5q\rangle|=2G_D|
\langle \overline{q}^c i\gamma_5\mathcal{T}_7\lambda_7q\rangle|.\quad
\end{eqnarray}
Hereafter the index $f$ is suppressed in order to stress the degeneracy of flavors. 
In the considered case
\begin{eqnarray}
\label{XXX_O}
\frac{T}{2V}{\rm Tr\ln(\beta\mathcal{S}^{-1})}&=&
T\sum_n\int\frac{d{\bf k}}{(2\pi)^3}\left[
8\left(\ln\frac{\omega_n^2+(\omega^+_{\rm oct})^2}{T^2}+\ln\frac{\omega_n^2+(\omega^-_{\rm oct})^2}{T^2}\right)\right.\nonumber\\
& &\hspace*{1cm}+\left.
\ln\frac{\omega_n^2+(\omega^+_{\rm sing})^2}{T^2}+\ln\frac{\omega_n^2+(\omega^-_{\rm sing})^2}{T^2}\right].\quad
\end{eqnarray}
Here $\omega_n=\pi T(2n+1)$ is a fermionic Matsubara frequency. Performing the summation over the corresponding index and taking the limit $T\rightarrow 0$, which is of interest for the astrophysical applications, we arrive at the zero temperature thermodynamic potential of the CFL phase of massless interacting quark matter
\begin{eqnarray}
\label{XXXI_O}
\Omega&=&\Omega_q-G_V\langle q^+q\rangle^2+\frac{3\Delta^2}{4G_D}+\mathcal{U}^{(0)}-\langle\overline{q}\Sigma^{(0)} q\rangle,
\end{eqnarray}
with
\begin{eqnarray}
\label{XXXII_O}
\Omega_q=-\int\frac{d{\bf k}}{(2\pi)^3}
\left[8\left(\omega^+_{\rm oct}+\omega^-_{\rm oct}\right)+\omega^+_{\rm sing}+\omega^-_{\rm sing}\right]
\end{eqnarray}
being a quark part. 
The momentum integrals in Eq. (\ref{XXXII_O}) are regularized by a sharp cutoff $\Lambda$. 
The definitions of the quark self-energy $\Sigma^{(0)}$ and the diquark gap $\Delta$ provide stationarity of the thermodynamic potential, which is equivalent to the gap equations 
\begin{eqnarray}
\label{XXXIII_O}
3\sigma&=&\frac{1}{D_0}\int\frac{d{\bf k}}{(2\pi)^3}
\left[8\left(\frac{\omega+\mu^*}{\omega^+_{\rm oct}}+\frac{\omega-\mu^*}{\omega^-_{\rm oct}}\right)+\frac{\omega+\mu^*}{\omega^+_{\rm sing}}+\frac{\omega-\mu^*}{\omega^-_{\rm sing}}\right]\frac{m^*}{\omega},\nonumber\\
\\
\label{XXXIV_O}
3\Delta&=&4G_D\Delta\int\frac{d{\bf k}}{(2\pi)^3}
\left[8\left(\frac{1}{\omega^+_{\rm oct}}+\frac{1}{\omega^-_{\rm oct}}\right)+4\left(\frac{1}{\omega^+_{\rm sing}}+\frac{1}{\omega^-_{\rm sing}}\right)\right].
\end{eqnarray}
Let us first analyze the equation for the reduced mass parameter. Since $m^*\propto\sigma$, then it obviously has a trivial solution representing the chirally restored CFL phase. For another one the reduced mass parameter has its vacuum value. In this case effective quark mass diverges and the integrand in Eq. (\ref{XXXIII_O}) simplifies to $18$. This solution corresponds to the chirally broken normal phase. Thus
\begin{eqnarray}
\label{XXXV_O}
\sigma_{\rm \chi SB}&=&\sigma_0~,~~
\sigma_{\rm CFL}=0.
\label{XXXVI_O}
\end{eqnarray}
Note that $\sigma_0=-{\langle\overline{f}f\rangle_0}/{D_0}$ is connected to the vacuum value of a single flavor chiral condensate $\langle\overline{f}f\rangle_0$ which defines the momentum cutoff by the relation
\begin{eqnarray}
\label{XXXVII_O}
\sigma_0=\frac{6}{D_0}\int\frac{d{\bf k}}{(2\pi)^3}=\frac{\Lambda^3}{\pi^2 D_0}.
\end{eqnarray}
We used $\langle\overline{f}f\rangle_0=-(251~{\rm MeV})^3$, which yields $\Lambda=538~{\rm MeV}$. 
This value of the momentum cutoff is used in order to parametrize vector and diquark couplings by $\eta_V=G_V\Lambda^2$ and $\eta_D=G_D\Lambda^2$. 
The models considered below are labelled by pairs of numbers $(\eta_V,\eta_D)$. For example, $(1.0,2.0)$ corresponds to $G_V=\Lambda^{-2}$ and $G_D=2\Lambda^{-2}$.

The effective quark mass and, consequently, $\omega^\pm_{\rm oct}$ and $\omega^\pm_{\rm sing}$ diverge in the phase with broken chiral symmetry, leading to a vanishing diquark pairing gap. 
In the CFL phase quarks are massless and this pairing gap becomes finite. Thus
\begin{eqnarray}
\label{XXXVIII_O}
\Delta_{\chi SB}&=&0,\\
\label{XXXIX_O}
3&=&\int\frac{d{\bf k}}{(2\pi)^3}
\left[\frac{32G_D}{\omega^+_{\rm oct}}+\frac{32G_D}{\omega^-_{\rm oct}}+\frac{16G_D}{\omega^+_{\rm sing}}+\frac{16G_D}{\omega^-_{\rm sing}}\right]_{m^*=0,~\Delta=\Delta_{\rm CFL}}.
\end{eqnarray}
\begin{figure}[t]
\includegraphics[width=0.47\columnwidth]{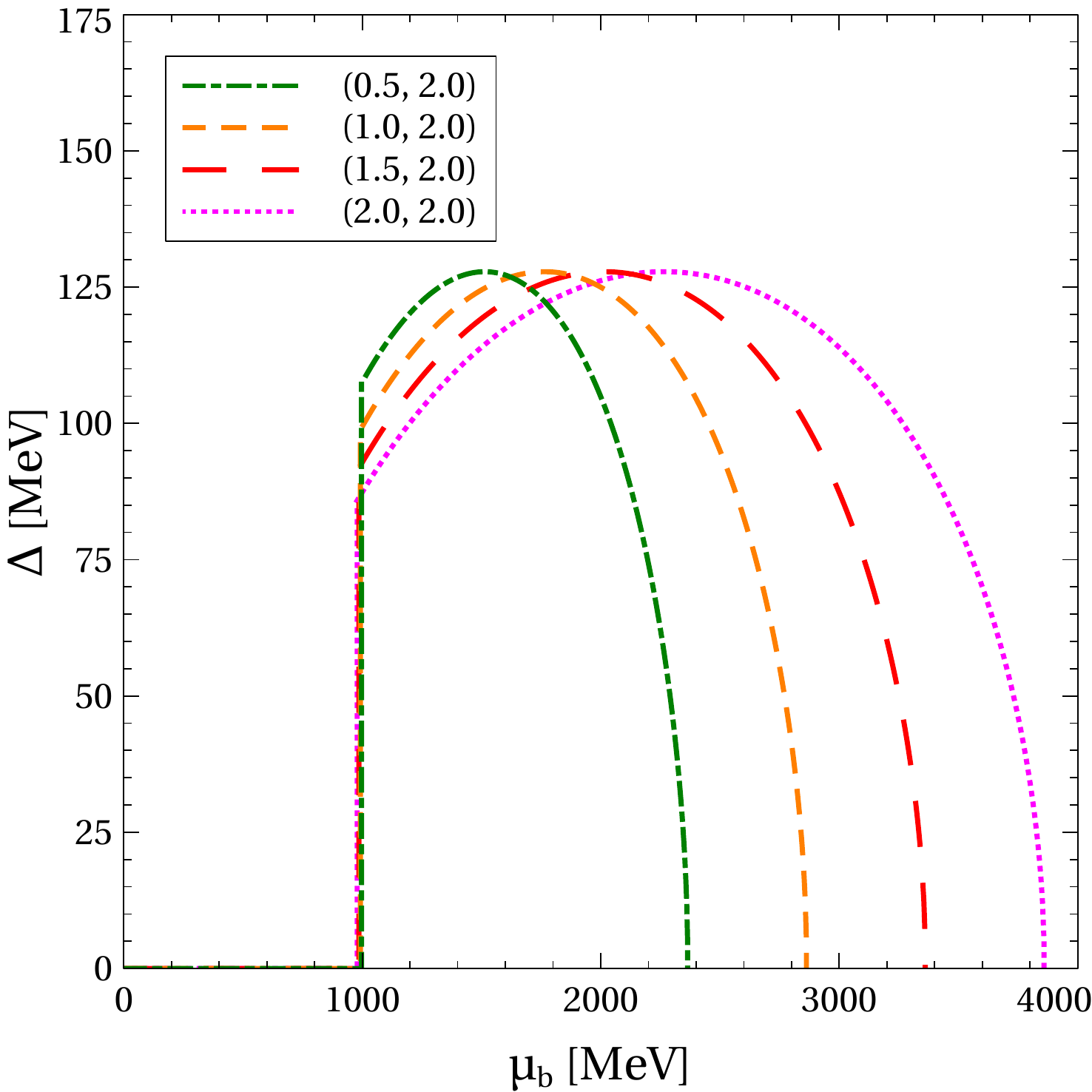}
\hfill
\includegraphics[width=0.47\columnwidth]{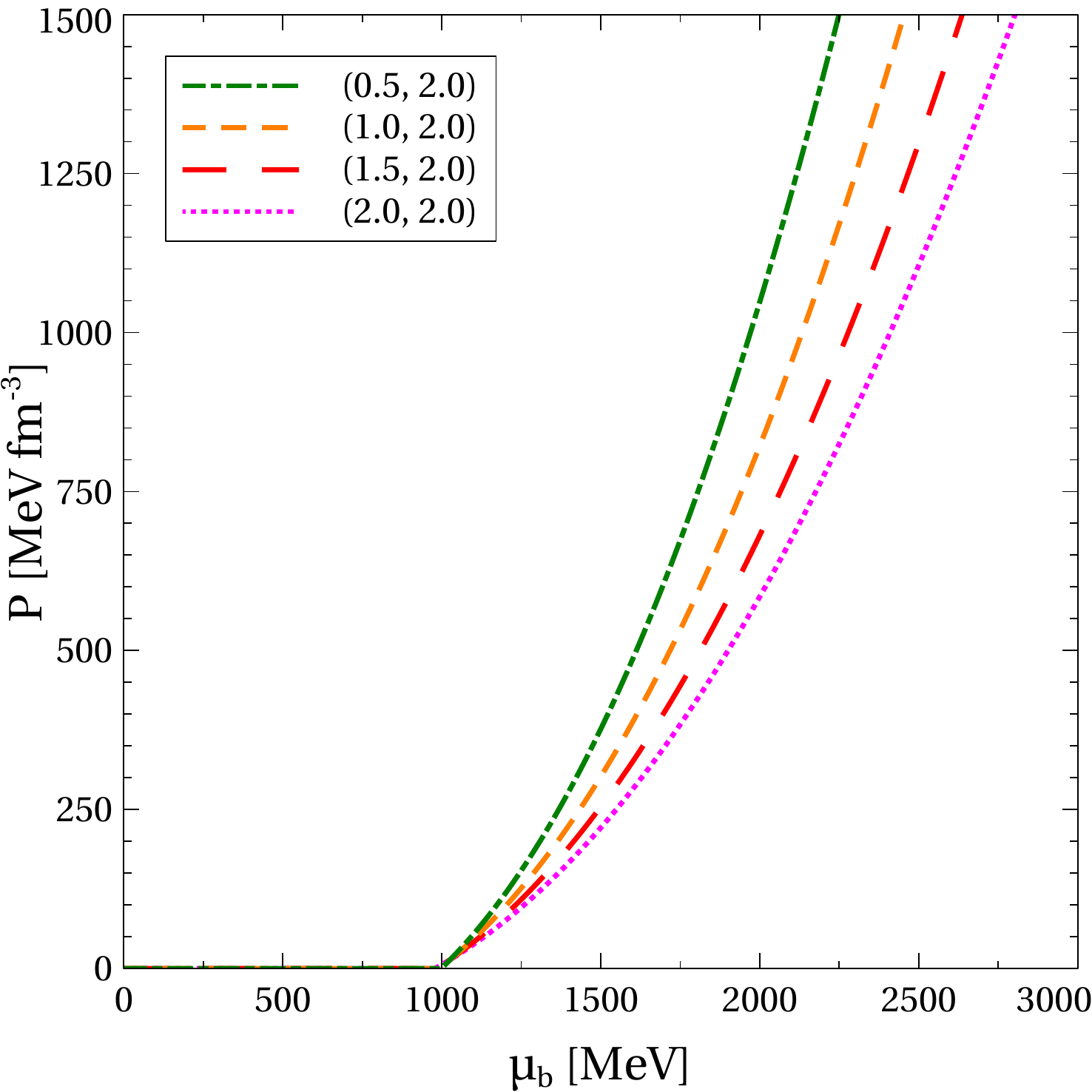}
\caption{Diquark pairing gap $\Delta$ (left panel) and pressure $P$ (right panel) of color superconducting quark matter as functions of baryonic chemical potential $\mu_b$.}
\label{fig1_O}
\end{figure}

The divergence of $m^*$ in the chirally broken phase leads to $\Omega_q=\langle\overline{q}\Sigma^{(0)}q\rangle=-3D_0\sigma m^*$, while $\mathcal{U}^{(0)}=0$ due to $\sigma=\sigma_0$. 
The quark number density and the diquark gap also vanish in this phase. Therefore, its pressure $P$ being the negative of the thermodynamic potential vanishes. 
In the CFL phase quarks are massless, which is equivalent to $\Sigma^{(0)}=0$. 
Therefore, the pressure of two phases
\begin{eqnarray}
\label{XXXX_O}
P_{\rm \chi SB}&=&0,\\
\label{XXXXI_O}
P_{\rm CFL}&=&-\Omega_q|_{m^*=0,~\Delta=\Delta_{\rm CFL}}+G_V\langle q^+q\rangle^2-\frac{3\Delta_{\rm CFL}^2}{4G_D}-\mathcal{U}^{(0)}_{\sigma=0}.
\end{eqnarray}
It is worth mentioning that the last term in the previous expression is constant and negative. 
Therefore, it can be related to the effective bag pressure 
\begin{eqnarray}
\label{XXXXII_O}
\mathcal{B}\equiv\mathcal{U}^{(0)}_{\sigma=0}=D_0^{\frac{5}{3}}(5\sigma_0^2)^{\frac{1}{3}}.
\end{eqnarray}
The quark number density of the CFL phase should be self-consistently found according to
\begin{eqnarray}
\label{XXXXIII_O}
\langle q^+q\rangle_{\rm CFL}=-\frac{\partial \Omega_q}{\partial\mu}\Bigl|_{m^*=0,~\Delta=\Delta_{\rm CFL}}.
\end{eqnarray}
This number density is related to the baryonic charge density $n_b$ as $\langle q^+ q\rangle =3n_b$, while the chemical potential of the baryonic charge is $\mu_b=3\mu$. 
It is also important to note that equality of masses and chemical potentials for the three quark flavors leads to the equality of their partial number densities. As a result, the quark matter is electrically and color neutral.

\begin{figure}[t]
\includegraphics[width=\columnwidth]{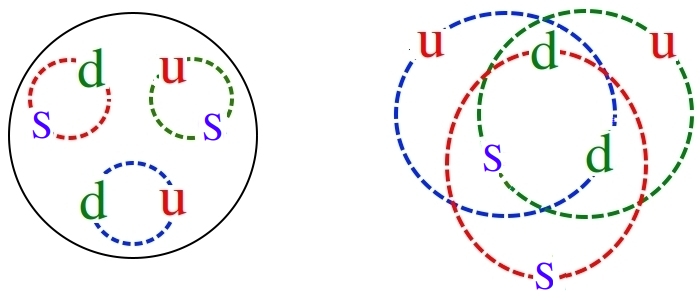}
\caption{Illustration of the transition from a BEC of sexaquarks as 3-diquark bound state (left panel) to a BCS condensate of diquarks in the CFL phase (right panel).}
\label{figBEC_BCS}
\end{figure}

Two solutions of the present model should be merged at $\mu_b$ providing $P_{\chi SB}=P_{\rm CFL}$. 
At low chemical potentials the solution corresponding to the CFL phase has negative pressure $P_{\rm CFL}<0$ signalling that superconducting chirally symmetric quark matter is disfavoured in that region. 
At a certain $\mu_b$ the pressure $P_{\rm CFL}$ attains a zero value and gets positive at larger chemical potentials. 
That region corresponds to the chirally symmetric CFL phase of quark matter.

The dependence of the diquark pairing gap on the baryonic charge chemical potential is shown in the left panel of Fig.~\ref{fig1_O}. 
With this solution we evaluate the present model pressure as a function of baryonic chemical potential shown on the right panel of Fig.~\ref{fig1_O}.

We utilize the present model in order to construct the hybrid quark-hadron EoS by means of the Maxwell construction. For the hadron part we use the DD2Y-T+S EoS discussed in the previous section. As it was mentioned, the curve of this EoS terminates in the $\mu_b-P$ plane when chemical potential reaches the half of the sexaquark mass $m_S$. 
This manifests the onset of BEC of sexaquarks. 

While having a vanishing incompressibility $\mathcal{K}=9\partial P/\partial n_b$, this condensate is dynamically unstable against gravitational compression. 
A sufficient increase of the condensate density leads to the Mott dissociation of sexaquarks into three diquarks. 
This opens a window for the transition from the BEC of sexaquarks to the BCS phase with diquark condensation in CFL phase quark matter (see Fig. \ref{figBEC_BCS}). 
The condition of switching between these two regimes of strongly interacting matter is given by the Gibbs criterion of phase equilibrium formulated as $\mu_b={m_S}/{2}$, i.e. 
\begin{eqnarray}
P_{\rm hadron}(\mu_b={m_S}/{2})&=&
P_{\rm CFL}({\mu_b={m_S}/{2}})~.
\label{eq:phad}
\end{eqnarray}
We adjust the coupling $D_0$ in order to provide this condition. Below the transition point, the matter exists in the hadronic phase, while above the transition the matter is converted to the CFL phase of quark matter. 

\begin{figure}[t]
\includegraphics[width=0.47\columnwidth]{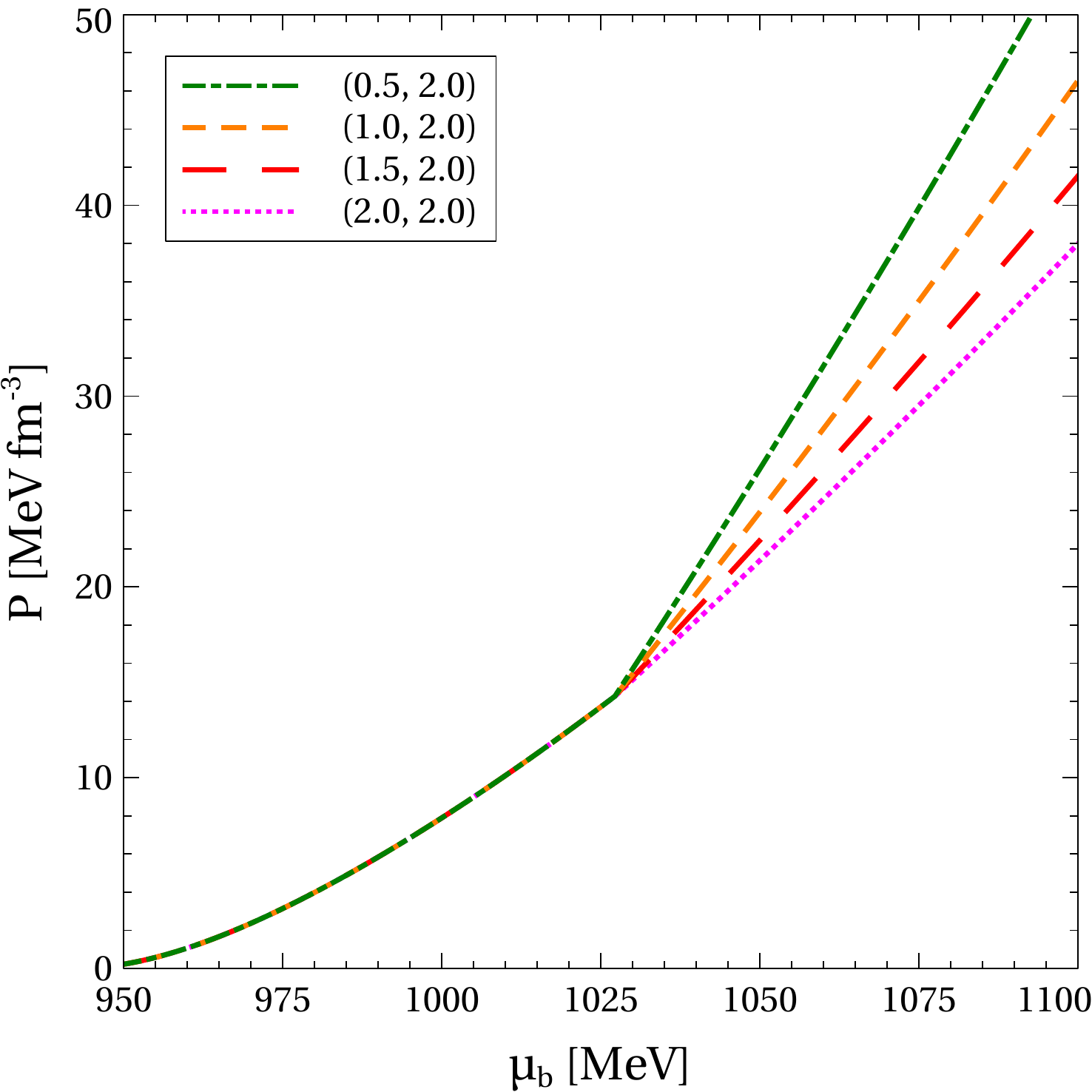}
\hfill
\includegraphics[width=0.47\columnwidth]{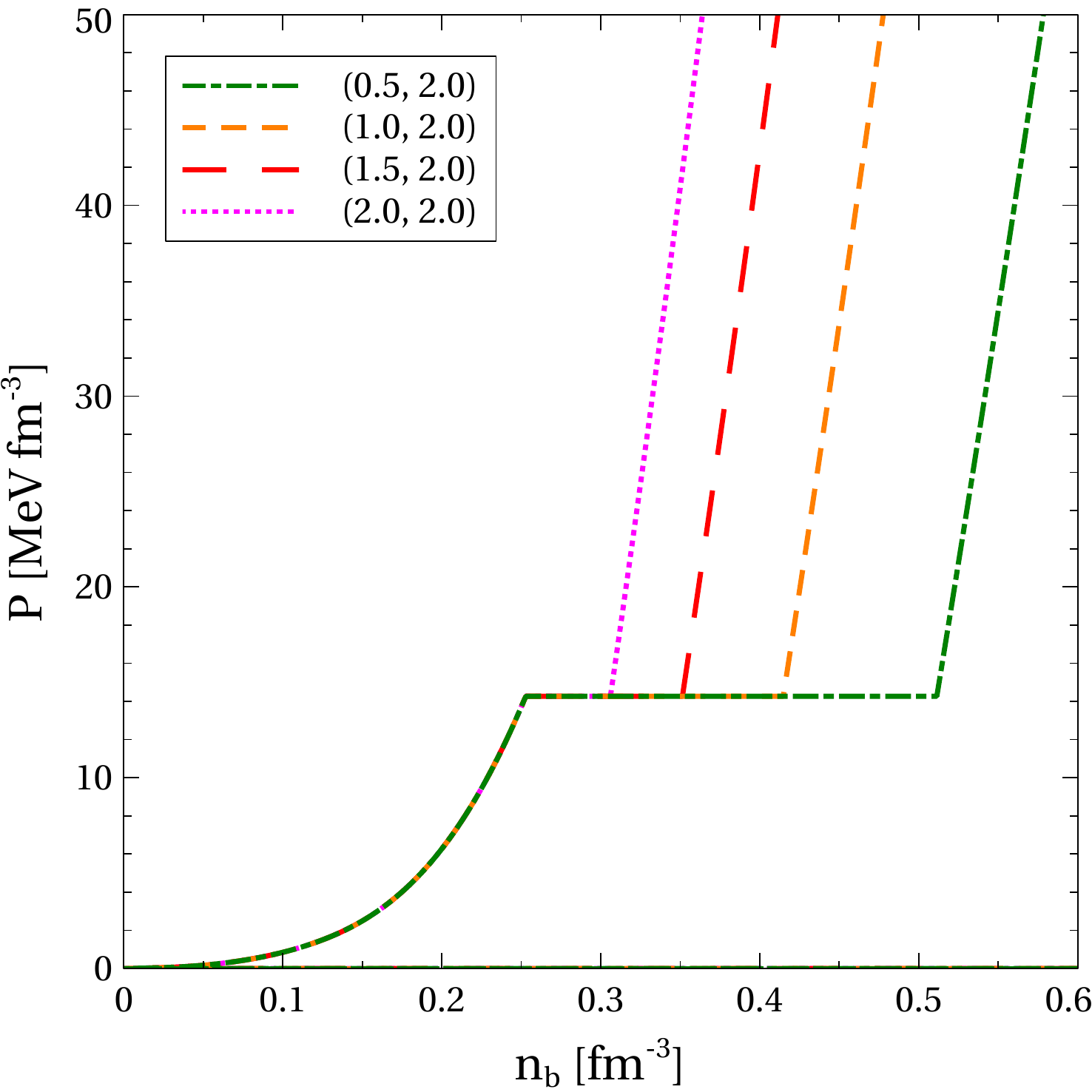}
\caption{Pressure $P$ of hybrid quark-hadron matter as function of the baryonic chemical potential $\mu_b$ (left panel) and as function of the baryon density $n_b$ (right panel).}
\label{fig2_O}
\end{figure}
\begin{figure}[t]
\centerline{
\includegraphics[width=0.7\columnwidth]{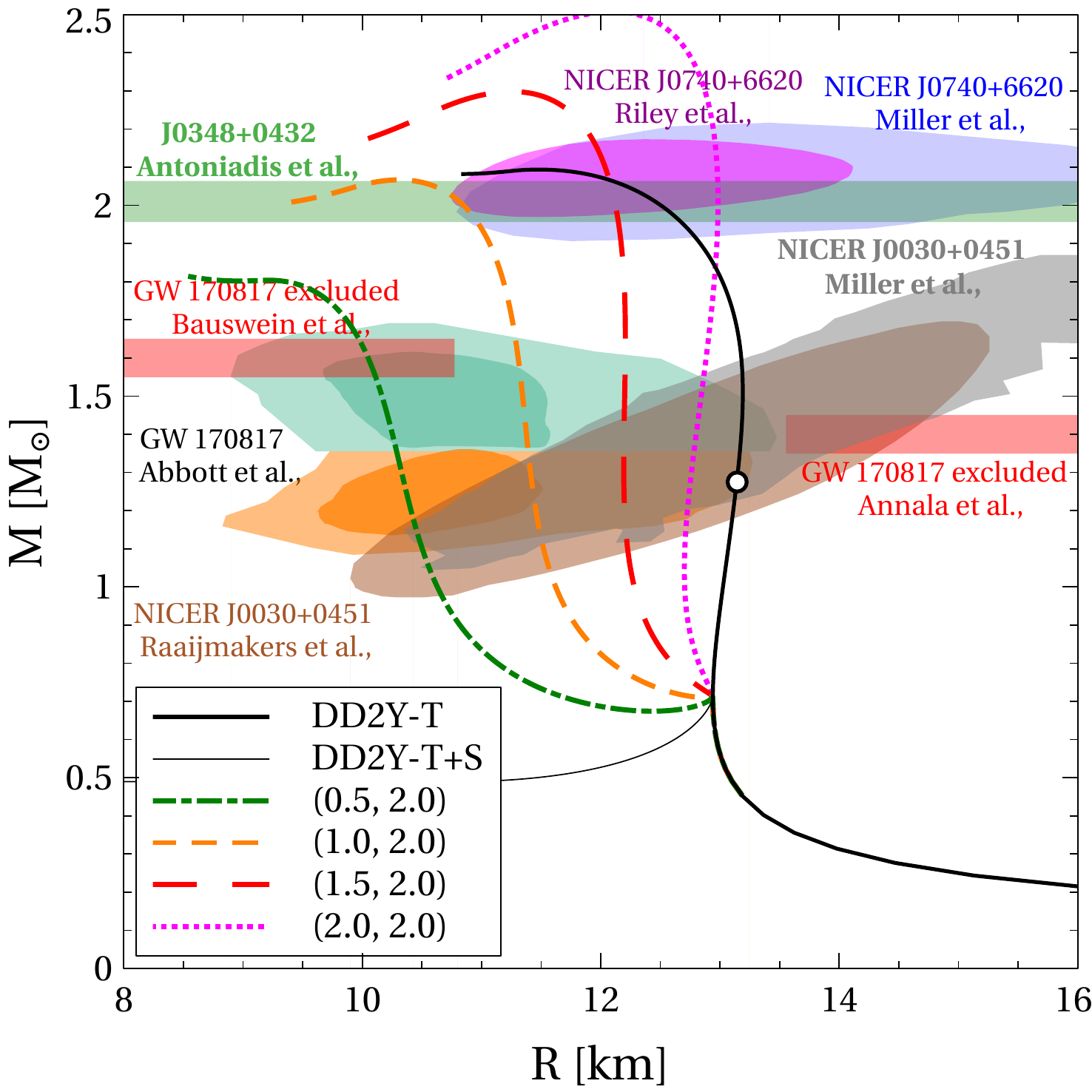}}
\caption{Mass-radius relation of hybrid star with the quark-hadron EoS presented on Fig. \ref{fig2_O}. The empty circle on the hadronic curve indicates the hyperon onset. 
Quark and hadron branches split at the sexaquark onset mass of about $0.7~M_\odot$. The astrophysical constraints depicted by the colored bends and shaded areas are discussed in the text.}
\label{fig3_O}
\end{figure}

The hybrid quark-hadron EoS is shown on Fig. \ref{fig2_O}. 
As is seen from the left panel, location of the transition point in the $\mu_b-P$ plane is not affected by the values of the vector and diquark couplings of quark matter provided by a proper choice of $D_0$. 
The right panel of Fig. \ref{fig2_O} demonstrates that the limiting baryon density of hadron matter also remains unchanged under the variation of $\eta_V$ and $\eta_D$. 
At the same time, the baryonic density exhibits a discontinuous jump along the transition point signalling the first order phase transition. The onset density of quark matter decreases with the growth of the vector coupling. 

A larger $\eta_V$ also leads to a stiffening of the quark matter branch of the hybrid EoS. 
This is reflected in the increase of the maximum mass of compact stars with quark cores. 
The corresponding mass-radius diagram is shown on Fig. \ref{fig3_O}. 
Quark and hadron branches split at the mass of star for which in the center the conditions for the appearance of the sexaquark and thus its BEC are met.
The purely hadronic solution (thin black curve) becomes unstable after that, thus limiting the maximum stellar mass to about $0.7~M_\odot$. 
Switching to the quark branches stiffens the EoS providing stability of the stellar matter against the gravitational collapse. 
As a result the quark branches of the solution (colored curves) lead to much larger maximum stellar masses, which increase with the value of the vector coupling. 
For $\eta_D=2.0$, the maximum mass reaches the two solar mass limit at $\eta_V$ slightly below one. 
As can be seen in Fig. \ref{fig3_O}, the present model is in a reasonable agreement with the astrophysical constraints discussed in the previous section and presented by the colored bands. 

\begin{figure}[t]
\includegraphics[width=0.47\columnwidth]{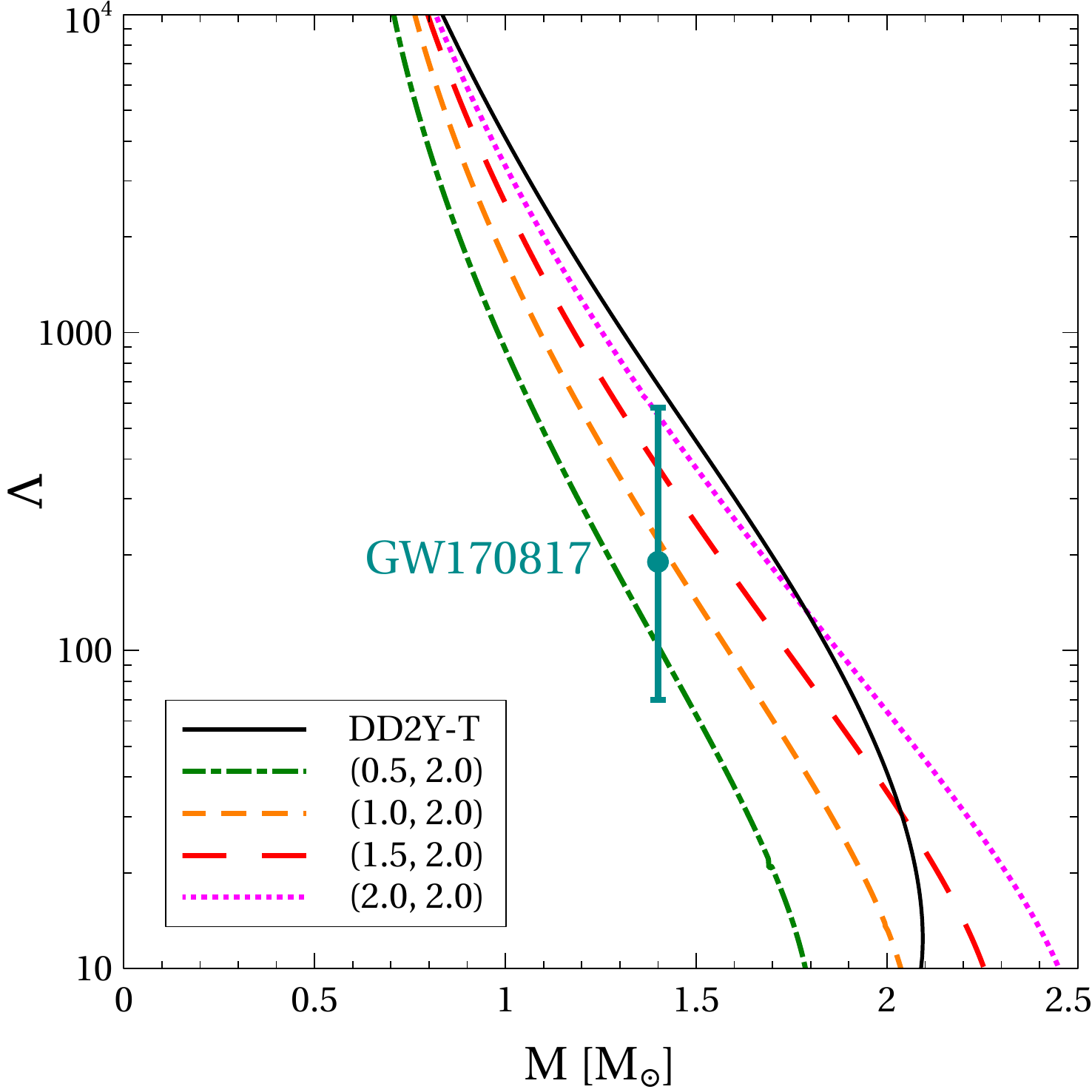}
\includegraphics[width=0.47\columnwidth]{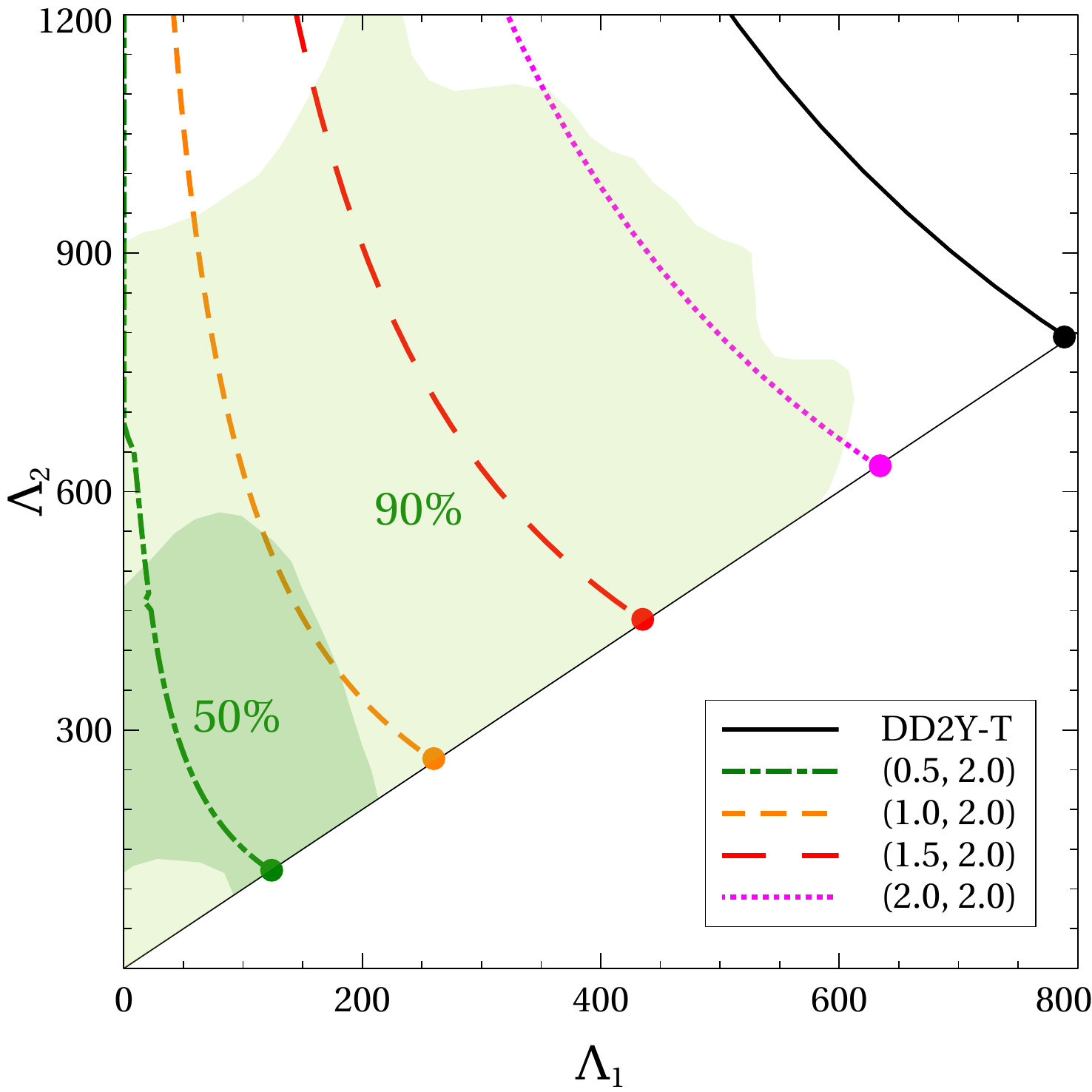}
\caption{Tidal deformability $\Lambda$ as a function of stellar mass $\rm M$ (left panel) and tidal deformability of the low mass component of the compact star merger $\Lambda_2$ as a function of the corresponding parameter $\Lambda_1$ of the high mass one found for the chirp mass $\mathcal{M}=1.188~M_\odot$ (right panel). The calculations are performed for hybrid stars with the quark-hadron EoS presented on Fig. \ref{fig2_O} (color curves) and compared to the results obtained for purely hadronic star with the DD2Y-T EoS (black curves).  Dark and light green shaded areas on the right panel demonstrate the regions falling into the $50~\%$ and $90~\%$ confidence levels, while filled circles represent the configurations with equal masses of two components $\rm M_1=M_2=1.3646~ M_\odot$.}
\label{fig4_O}
\end{figure}

Tidal deformability $\Lambda$ is another important characteristic of compact stars quantifying its response to an external gravitational field \cite{Peters:1963ux,LIGOScientific:2017vwq,LIGOScientific:2018cki}. This $\Lambda$ is fully determined by the properties of stellar matter and, therefore, can be found for a given mass of the star $M$ if its  EoS is known. Fig. \ref{fig4_O} shows dependence of the tidal deformability of compact quark-hadron star on its mass. The general trend is that $\Lambda$ decreases with $\rm M$ since heavier stars are more compact and, therefore, are less deformable by external gravitational field. Analysis of the gravitational wave signal GW 170817 implies that tidal deformability of the $1.4\rm M_\odot$ compact stars does not exceed 580, i.e. $\Lambda_{1.4}\le 580$ \cite{LIGOScientific:2018cki}. At $\eta_D=2.0$ this constraint limits $\eta_V$ from above by the value slightly above two. Note, the DD2Y-T EoS does not fit this constraint, while DD2Y-T+S EoS does not even reach $1.4~\rm M_\odot$ compact stars. 

The gravitational wave signal GW 170817 was produced by the merger of two compact objects with masses ${\rm M_1\ge M_2}$. Analysis of the observational data does not give the values of these masses but implies that they are related through the chirp mass \cite{Peters:1963ux}
\begin{eqnarray}
\label{XXXXV_O}
\mathcal{M}=\frac{(\rm M_1 M_2)^{\frac{3}{5}}}{(\rm M_1+M_2)^{\frac{1}{5}}}.
\end{eqnarray}
The best measured combination of the masses from GW 170817 yields $\mathcal{M}=1.188^{+0.004}_{-0.002}~M_\odot$ \cite{LIGOScientific:2017vwq}, which is used below. For each $\rm M_1$ and $M_2$ providing this value of the chirp mass we calculated tidal deformabilities $\Lambda_1$ and $\Lambda_2$. Relation between them is shown on the right panel of Fig. \ref{fig4_O}, which also depicts the areas corresponding to 50 \% and 90 \% confidence intervals \cite{LIGOScientific:2018cki}. At $\eta_D=2.0$ agreement with the observational data at the 90 \% confidence level is provided by $\eta_V\le2.0$. Purely hadronic DD2Y-T EoS lies well beyond the corresponding interval. Agreement at the 50 \% confidence level requires $\eta_V\le1.0$, which, however, contradicts to the constraints on the mass-radius relation. 

\section{Conclusions}

In this contribution, we present for the first time a scenario according to which early quark deconfinement in compact stars is triggered by the BEC of a light sexaquark ($m_S<2054$ MeV) that has been suggested as a  candidate particle to explain the baryonic DM in the Universe. 
The BEC onset of S marks the maximum mass of hadronic neutron stars and it occurs when the condition for the baryon chemical potential $\mu_b=m_S/2$ is fulfilled in the center of the star, corresponding to $M_{\rm onset}<0.7 M_\odot$. 
In the gravitational field of the star the density of BEC of S increases until a new state of the matter is attained which consists of dissociated S states, the CFL phase with a diquark condensate, thus presenting a form of BEC-BCS transition in strongly interacting matter.
For the description of the CFL phase, we have developed here for the first time the three-flavor extension of the density-functional formulation of a chirally symmetric Lagrangian model of quark matter with confining properties encoded in a divergence of the scalar selfenergy at low densities and temperatures. 
As this density functional model does not show sequential but rather simultaneous deconfinement of quark flavors, the "No-Go" theorem of thee NJL model against the possibility of absolutely stable strange quark matter does not apply here.  
However, since we had to make several approximations in this first evaluation of the three-flavor version of the density-functional model, we leave a discussion of the possibility of absolutely stable strange quark matter to a subsequent work.

\subsection*{Acknowledgements}
This work was supported by the Polish National Science Centre (NCN) under grant No. 2019/33/B/ST9/03059.
It was performed within a project that has received funding from the European Union's Horizon 2020 research and innovation program under grant agreement STRONG - 2020 - No. 824093.


\begin{thebibliography}{61}
\newcommand{\enquote}[1]{#1}
\providecommand{\natexlab}[1]{#1}
\providecommand{\url}[1]{\texttt{#1}}
\providecommand{\urlprefix}{URL }
\providecommand{\eprint}{eprint }
\expandafter\ifx\csname urlstyle\endcsname\relax
  \providecommand{\doi}[1]{doi:\discretionary{}{}{}#1}\else
  \providecommand{\doi}{doi:\discretionary{}{}{}\begingroup
  \urlstyle{rm}\Url}\fi

\bibitem[{Abbott \emph{et~al.}(2017)}]{LIGOScientific:2017vwq}
Abbott, B.~P. \emph{et~al.} (2017). \enquote{{GW170817: Observation of
  Gravitational Waves from a Binary Neutron Star Inspiral},} \emph{Phys. Rev.
  Lett.} \textbf{119}, 16, p. 161101, \doi{10.1103/PhysRevLett.119.161101},
  \href{http://arxiv.org/abs/1710.05832}{\UrlFont{arXiv:1710.05832 [gr-qc]}}.

\bibitem[{Abbott \emph{et~al.}(2018)}]{LIGOScientific:2018cki}
Abbott, B.~P. \emph{et~al.} (2018). \enquote{{GW170817: Measurements of neutron
  star radii and equation of state},} \emph{Phys. Rev. Lett.} \textbf{121}, 16,
  p. 161101, \doi{10.1103/PhysRevLett.121.161101},
  \href{http://arxiv.org/abs/1805.11581}{\UrlFont{arXiv:1805.11581 [gr-qc]}}.

\bibitem[{Ade \emph{et~al.}(2016)}]{Planck:2015fie}
Ade, P. A.~R. \emph{et~al.} (2016). \enquote{{Planck 2015 results. XIII.
  Cosmological parameters},} \emph{Astron. Astrophys.} \textbf{594}, p. A13,
  \doi{10.1051/0004-6361/201525830}.

\bibitem[{Alcock \emph{et~al.}(1986)Alcock, Farhi and Olinto}]{Alcock:1986hz}
Alcock, C., Farhi, E.,  and Olinto, A. (1986). \enquote{{Strange stars},}
  \emph{Astrophys. J.} \textbf{310}, pp. 261--272, \doi{10.1086/164679}.

\bibitem[{Alford \emph{et~al.}(2005)Alford, Braby, Paris and
  Reddy}]{Alford:2004pf}
Alford, M., Braby, M., Paris, M.~W.,  and Reddy, S. (2005). \enquote{{Hybrid
  stars that masquerade as neutron stars},} \emph{Astrophys. J.} \textbf{629},
  pp. 969--978, \doi{10.1086/430902},
  \href{http://arxiv.org/abs/nucl-th/0411016}{\UrlFont{arXiv:nucl-th/0411016}}.

\bibitem[{Annala \emph{et~al.}(2018)Annala, Gorda, Kurkela and
  Vuorinen}]{Annala:2017llu}
Annala, E., Gorda, T., Kurkela, A.,  and Vuorinen, A. (2018).
  \enquote{{Gravitational-wave constraints on the neutron-star-matter Equation
  of State},} \emph{Phys. Rev. Lett.} \textbf{120}, 17, p. 172703,
  \doi{10.1103/PhysRevLett.120.172703},
  \href{http://arxiv.org/abs/1711.02644}{\UrlFont{arXiv:1711.02644
  [astro-ph.HE]}}.

\bibitem[{Anti\'c \emph{et~al.}(2021)Anti\'c, Shahrbaf, Blaschke and
  Grunfeld}]{Antic:2021zbn}
Anti\'c, S., Shahrbaf, M., Blaschke, D.,  and Grunfeld, A.~G. (2021).
  \enquote{{Parameter mapping between the microscopic nonlocal
  Nambu-Jona-Lasinio and constant-sound-speed models},}
  \href{http://arxiv.org/abs/arXiv:2105.00029}{\UrlFont{arXiv:arXiv:2105.00029
  [nucl-th]}}.

\bibitem[{Antoniadis \emph{et~al.}(2013)}]{Antoniadis:2013pzd}
Antoniadis, J. \emph{et~al.} (2013). \enquote{{A Massive Pulsar in a Compact
  Relativistic Binary},} \emph{Science} \textbf{340}, p. 6131,
  \doi{10.1126/science.1233232},
  \href{http://arxiv.org/abs/1304.6875}{\UrlFont{arXiv:1304.6875
  [astro-ph.HE]}}.

\bibitem[{Baldo \emph{et~al.}(2003)Baldo, Burgio and Schulze}]{Baldo:2003vx}
Baldo, M., Burgio, G.~F.,  and Schulze, H.~J. (2003). \enquote{{Neutron star
  structure with hyperons and quarks},} in \emph{{NATO Advanced Research
  Workshop on Superdense QCD Matter and Compact Stars}}, pp. 113--134,
  \href{http://arxiv.org/abs/astro-ph/0312446}{\UrlFont{arXiv:astro-ph/0312446}}.

\bibitem[{Bauswein \emph{et~al.}(2017)Bauswein, Just, Janka and
  Stergioulas}]{Bauswein:2017vtn}
Bauswein, A., Just, O., Janka, H.-T.,  and Stergioulas, N. (2017).
  \enquote{{Neutron-star radius constraints from GW170817 and future
  detections},} \emph{Astrophys. J. Lett.} \textbf{850}, 2, p. L34,
  \doi{10.3847/2041-8213/aa9994},
  \href{http://arxiv.org/abs/1710.06843}{\UrlFont{arXiv:1710.06843
  [astro-ph.HE]}}.

\bibitem[{Baym and Chin(1976)}]{Baym:1976yu}
Baym, G. and Chin, S.~A. (1976). \enquote{{Can a Neutron Star Be a Giant MIT
  Bag?}} \emph{Phys. Lett. B} \textbf{62}, pp. 241--244,
  \doi{10.1016/0370-2693(76)90517-7}.

\bibitem[{Bazavov \emph{et~al.}(2019)}]{HotQCD:2018pds}
Bazavov, A. \emph{et~al.} (2019). \enquote{{Chiral crossover in QCD at zero and
  non-zero chemical potentials},} \emph{Phys. Lett. B} \textbf{795}, pp.
  15--21, \doi{10.1016/j.physletb.2019.05.013},
  \href{http://arxiv.org/abs/1812.08235}{\UrlFont{arXiv:1812.08235 [hep-lat]}}.

\bibitem[{Blaschke \emph{et~al.}(2010)Blaschke, Kl\"ahn, Lastowiecki and
  Sandin}]{Blaschke:2010vd}
Blaschke, D., Kl\"ahn, T., Lastowiecki, R.,  and Sandin, F. (2010). \enquote{{How
  strange are compact star interiors ?}} \emph{J. Phys. G} \textbf{37}, p.
  094063, \doi{10.1088/0954-3899/37/9/094063},
  \href{http://arxiv.org/abs/1002.1299}{\UrlFont{arXiv:1002.1299 [nucl-th]}}.

\bibitem[{Blaschke \emph{et~al.}(2009)Blaschke, Sandin, Kl\"ahn and
  Berdermann}]{Blaschke:2008br}
Blaschke, D., Sandin, F., Kl\"ahn, T.,  and Berdermann, J. (2009).
  \enquote{{Sequential deconfinement of quark flavors in neutron stars},}
  \emph{Phys. Rev. C} \textbf{80}, p. 065807, \doi{10.1103/PhysRevC.80.065807},
  \href{http://arxiv.org/abs/0807.0414}{\UrlFont{arXiv:0807.0414 [nucl-th]}}.

\bibitem[{Bodmer(1971)}]{Bodmer:1971we}
Bodmer, A.~R. (1971). \enquote{{Collapsed nuclei},} \emph{Phys. Rev. D}
  \textbf{4}, pp. 1601--1606, \doi{10.1103/PhysRevD.4.1601}.


\bibitem[{Burgio \emph{et~al.}(2002)Burgio, Baldo, Sahu and
  Schulze}]{Burgio:2002sn}
Burgio, G.~F., Baldo, M., Sahu, P.~K.,  and Schulze, H.~J. (2002).
  \enquote{{The Hadron quark phase transition in dense matter and neutron
  stars},} \emph{Phys. Rev. C} \textbf{66}, p. 025802,
  \doi{10.1103/PhysRevC.66.025802},
  \href{http://arxiv.org/abs/nucl-th/0206009}{\UrlFont{arXiv:nucl-th/0206009}}.

\bibitem[{Chodos \emph{et~al.}(1974{\natexlab{a}})Chodos, Jaffe, Johnson and
  Thorn}]{Chodos:1974pn}
Chodos, A., Jaffe, R.~L., Johnson, K.,  and Thorn, C.~B. (1974{\natexlab{a}}).
  \enquote{{Baryon Structure in the Bag Theory},} \emph{Phys. Rev. D}
  \textbf{10}, p. 2599, \doi{10.1103/PhysRevD.10.2599}.

\bibitem[{Chodos \emph{et~al.}(1974{\natexlab{b}})Chodos, Jaffe, Johnson, Thorn
  and Weisskopf}]{Chodos:1974je}
Chodos, A., Jaffe, R.~L., Johnson, K., Thorn, C.~B.,  and Weisskopf, V.~F.
  (1974{\natexlab{b}}). \enquote{{A New Extended Model of Hadrons},}
  \emph{Phys. Rev. D} \textbf{9}, pp. 3471--3495,
  \doi{10.1103/PhysRevD.9.3471}.

\bibitem[{Farhi and Jaffe(1984)}]{Farhi:1984qu}
Farhi, E. and Jaffe, R.~L. (1984). \enquote{{Strange Matter},} \emph{Phys. Rev.
  D} \textbf{30}, p. 2379, \doi{10.1103/PhysRevD.30.2379}.

\bibitem[{Farrar(2003)}]{Farrar:2002ic}
Farrar, G.~R. (2003). \enquote{{A stable H dibaryon: Dark matter candidate
  within QCD?}} \emph{Int. J. Theor. Phys.} \textbf{42}, pp. 1211--1218,
  \doi{10.1023/A:1025702431127}.

\bibitem[{Farrar(2017)}]{Farrar:2017eqq}
Farrar, G.~R. (2017). \enquote{{Stable Sexaquark},}
  \href{http://arxiv.org/abs/arXiv:1708.08951}{\UrlFont{arXiv:arXiv:1708.08951
  [hep-ph]}}.

\bibitem[{Farrar(2018{\natexlab{a}})}]{Farrar:2017ysn}
Farrar, G.~R. (2018{\natexlab{a}}). \enquote{{6-quark Dark Matter},} \emph{PoS}
  \textbf{ICRC2017}, p. 929, \doi{10.22323/1.301.0929},
  \href{http://arxiv.org/abs/arXiv:1711.10971}{\UrlFont{arXiv:arXiv:1711.10971
  [hep-ph]}}.

\bibitem[{Farrar(2018{\natexlab{b}})}]{Farrar:2018hac}
Farrar, G.~R. (2018{\natexlab{b}}). \enquote{{A precision test of the nature of
  Dark Matter and a probe of the QCD phase transition},}
  \href{http://arxiv.org/abs/arXiv:1805.03723}{\UrlFont{arXiv:arXiv:1805.03723
  [hep-ph]}}.

\bibitem[{Farrar \emph{et~al.}(2020)Farrar, Wang and Xu}]{Farrar:2020zeo}
Farrar, G.~R., Wang, Z.,  and Xu, X. (2020). \enquote{{Dark Matter Particle in
  QCD},}  \href{http://arxiv.org/abs/2007.10378}{\UrlFont{arXiv:2007.10378
  [hep-ph]}}.

\bibitem[{Farrar and Zaharijas(2003)}]{Farrar:2003gh}
Farrar, G.~R. and Zaharijas, G. (2003). \enquote{{Non-binding of flavor-singlet
  hadrons to nuclei},} \emph{Phys. Lett. B} \textbf{559}, pp. 223--228,
  \doi{10.1016/S0370-2693(03)00331-9},
  \href{http://arxiv.org/abs/hep-ph/0302190}{\UrlFont{arXiv:hep-ph/0302190}},
  [Erratum: Phys.Lett.B 575, 358--358 (2003)].

\bibitem[{Farrar and Zaharijas(2004)}]{Farrar:2003qy}
Farrar, G.~R. and Zaharijas, G. (2004). \enquote{{Nuclear and nucleon
  transitions of the H dibaryon},} \emph{Phys. Rev. D} \textbf{70}, p. 014008,
  \doi{10.1103/PhysRevD.70.014008}.

\bibitem[{Fonseca \emph{et~al.}(2021)}]{Fonseca:2021wxt}
Fonseca, E. \emph{et~al.} (2021). \enquote{{Refined Mass and Geometric
  Measurements of the High-Mass PSR J0740+6620},}
  \href{http://arxiv.org/abs/2104.00880}{\UrlFont{arXiv:2104.00880
  [astro-ph.HE]}}.

\bibitem[{Gerlach(1968)}]{Gerlach:1968zz}
Gerlach, U.~H. (1968). \enquote{{Equation of State at Supranuclear Densities
  and the Existence of a Third Family of Superdense Stars},} \emph{Phys. Rev.}
  \textbf{172}, pp. 1325--1330, \doi{10.1103/PhysRev.172.1325}.

\bibitem[{Gross \emph{et~al.}(2018)Gross, Polosa, Strumia, Urbano and
  Xue}]{Gross:2018ivp}
Gross, C., Polosa, A., Strumia, A., Urbano, A.,  and Xue, W. (2018).
  \enquote{{Dark Matter in the Standard Model?}} \emph{Phys. Rev. D}
  \textbf{98}, 6, p. 063005, \doi{10.1103/PhysRevD.98.063005},
  \href{http://arxiv.org/abs/1803.10242}{\UrlFont{arXiv:1803.10242 [hep-ph]}}.

\bibitem[{Haensel \emph{et~al.}(1986)Haensel, Zdunik and
  Schaeffer}]{Haensel:1986qb}
Haensel, P., Zdunik, J.~L.,  and Schaeffer, R. (1986). \enquote{{Strange quark
  stars},} \emph{Astron. Astrophys.} \textbf{160}, pp. 121--128.

\bibitem[{Hebeler \emph{et~al.}(2013)Hebeler, Lattimer, Pethick and
  Schwenk}]{Hebeler:2013nza}
Hebeler, K., Lattimer, J.~M., Pethick, C.~J.,  and Schwenk, A. (2013).
  \enquote{{Equation of state and neutron star properties constrained by
  nuclear physics and observation},} \emph{Astrophys. J.} \textbf{773}, p.~11,
  \doi{10.1088/0004-637X/773/1/11},
  \href{http://arxiv.org/abs/1303.4662}{\UrlFont{arXiv:1303.4662
  [astro-ph.SR]}}.

\bibitem[{Horowitz \emph{et~al.}(1985)Horowitz, Moniz and
  Negele}]{Horowitz:1985tx}
Horowitz, C.~J., Moniz, E.~J.,  and Negele, J.~W. (1985). \enquote{{HADRON
  STRUCTURE IN A SIMPLE MODEL OF QUARK / NUCLEAR MATTER},} \emph{Phys. Rev. D}
  \textbf{31}, pp. 1689--1699, \doi{10.1103/PhysRevD.31.1689}.

\bibitem[{Itoh(1970)}]{Itoh:1970uw}
Itoh, N. (1970). \enquote{{Hydrostatic Equilibrium of Hypothetical Quark
  Stars},} \emph{Prog. Theor. Phys.} \textbf{44}, p. 291,
  \doi{10.1143/PTP.44.291}.

\bibitem[{Ivanenko and Kurdgelaidze(1965)}]{Ivanenko:1965dg}
Ivanenko, D.~D. and Kurdgelaidze, D.~F. (1965). \enquote{{Hypothesis concerning
  quark stars},} \emph{Astrophysics} \textbf{1}, pp. 251--252,
  \doi{10.1007/BF01042830}.

\bibitem[{Ivanytskyi \emph{et~al.}(2021)Ivanytskyi, Blaschke and
  Maslov}]{Ivanytskyi:2021dgq}
Ivanytskyi, O., Blaschke, D.,  and Maslov, K. (2021). \enquote{{Confining
  density functional approach for color superconducting quark matter and
  mesonic correlations},}
  \href{http://arxiv.org/abs/2112.09223}{\UrlFont{arXiv:2112.09223 [nucl-th]}}.
  
\bibitem[{Ivanytskyi \emph{et~al.}(2022)Ivanytskyi and Blaschke}]{Ivanytskyi:2022oxv}
Ivanytskyi, O., and Blaschke, D. (2022). \enquote{{Density functional approach to quark matter with confinement and color superconductivity},}
  \href{https://arxiv.org/abs/2204.03611}{\UrlFont{arXiv:2204.03611 [nucl-th]}}.    

\bibitem[{Kaltenborn \emph{et~al.}(2017)Kaltenborn, Bastian and
  Blaschke}]{Kaltenborn:2017hus}
Kaltenborn, M. A.~R., Bastian, N.-U.~F.,  and Blaschke, D.~B. (2017).
  \enquote{{Quark-nuclear hybrid star equation of state with excluded volume
  effects},} \emph{Phys. Rev. D} \textbf{96}, 5, p. 056024,
  \doi{10.1103/PhysRevD.96.056024},
  \href{http://arxiv.org/abs/1701.04400}{\UrlFont{arXiv:1701.04400
  [astro-ph.HE]}}.

\bibitem[{Kl{\"a}hn \emph{et~al.}(2007)Kl{\"a}hn, Blaschke, Sandin, Fuchs,
  Faessler, Grigorian, R{\"o}pke and Tr{\"u}mper}]{Klahn:2006iw}
Kl{\"a}hn, T., Blaschke, D., Sandin, F., Fuchs, C., Faessler, A., Grigorian,
  H., R{\"o}pke, G.,  and Tr{\"u}mper, J. (2007). \enquote{{Modern compact star
  observations and the quark matter equation of state},} \emph{Phys. Lett. B}
  \textbf{654}, pp. 170--176, \doi{10.1016/j.physletb.2007.08.048},
  \href{http://arxiv.org/abs/nucl-th/0609067}{\UrlFont{arXiv:nucl-th/0609067}}.

\bibitem[{Kl\"ahn and Blaschke(2018)}]{Klahn:2017cyo}
Kl\"ahn, T. and Blaschke, D.~B. (2018). \enquote{{Strange matter in compact
  stars},} \emph{EPJ Web Conf.} \textbf{171}, p. 08001,
  \doi{10.1051/epjconf/201817108001},
  \href{http://arxiv.org/abs/1711.11260}{\UrlFont{arXiv:1711.11260 [hep-ph]}}.

\bibitem[{Kl\"ahn \emph{et~al.}(2006)}]{Klahn:2006ir}
Kl\"ahn, T. \emph{et~al.} (2006). \enquote{{Constraints on the high-density
  nuclear equation of state from the phenomenology of compact stars and
  heavy-ion collisions},} \emph{Phys. Rev. C} \textbf{74}, p. 035802,
  \doi{10.1103/PhysRevC.74.035802}.

\bibitem[{Kl\"ahn \emph{et al.}(2013)Kl\"ahn, Lastowiecki and Blaschke}]{Klahn:2013kga}
Kl\"ahn, T., Lastowiecki, R., and Blaschke, D.~B. (2013).
\enquote{{Implications of the measurement of pulsars with two solar masses
for quark matter in compact stars and heavy-ion collisions: A
Nambu-Jona-Lasinio model case study},} \emph{Phys. Rev.} \textbf{D88}, 8,
p. 085001, \doi{10.1103/PhysRevD.88.085001},
\href{http://arxiv.org/abs/1307.6996}{\UrlFont{arXiv:1307.6996 [nucl-th]}}.

\bibitem[{Marques \emph{et~al.}(2017)Marques, Oertel, Hempel and
  Novak}]{Marques:2017zju}
Marques, M., Oertel, M., Hempel, M.,  and Novak, J. (2017). \enquote{{New
  temperature dependent hyperonic equation of state: Application to rotating
  neutron star models and $I\text{-}Q$ relations},} \emph{Phys. Rev. C}
  \textbf{96}, 4, p. 045806, \doi{10.1103/PhysRevC.96.045806}.

\bibitem[{Miller \emph{et~al.}(2020)Miller, Chirenti and Lamb}]{Miller:2019nzo}
Miller, M.~C., Chirenti, C.,  and Lamb, F.~K. (2020). \enquote{{Constraining
  the equation of state of high-density cold matter using nuclear and
  astronomical measurements},} \emph{Astrophys. J.} \textbf{888}, p.~12,
  \doi{10.3847/1538-4357/ab4ef9},
  \href{http://arxiv.org/abs/1904.08907}{\UrlFont{arXiv:1904.08907
  [astro-ph.HE]}}.

\bibitem[{Miller \emph{et~al.}(2019)}]{Miller:2019cac}
Miller, M.~C. \emph{et~al.} (2019). \enquote{{PSR J0030+0451 Mass and Radius
  from $NICER$ Data and Implications for the Properties of Neutron Star
  Matter},} \emph{Astrophys. J. Lett.} \textbf{887}, 1, p. L24,
  \doi{10.3847/2041-8213/ab50c5}.

\bibitem[{Miller \emph{et~al.}(2021)}]{Miller:2021qha}
Miller, M.~C. \emph{et~al.} (2021). \enquote{{The Radius of PSR J0740+6620 from
  NICER and XMM-Newton Data},} \emph{Astrophys. J. Lett.} \textbf{918}, 2, p.
  L28, \doi{10.3847/2041-8213/ac089b},
  \href{http://arxiv.org/abs/2105.06979}{\UrlFont{arXiv:2105.06979
  [astro-ph.HE]}}.

\bibitem[{Oertel \emph{et~al.}(2017)Oertel, Hempel, Kl\"ahn and
  Typel}]{Oertel:2016bki}
Oertel, M., Hempel, M., Kl\"ahn, T.,  and Typel, S. (2017). \enquote{{Equations
  of state for supernovae and compact stars},} \emph{Rev. Mod. Phys.}
  \textbf{89}, 1, p. 015007, \doi{10.1103/RevModPhys.89.015007}.

\bibitem[{Peters and Mathews(1963)}]{Peters:1963ux}
Peters, P.~C. and Mathews, J. (1963). \enquote{{Gravitational radiation from
  point masses in a Keplerian orbit},} \emph{Phys. Rev.} \textbf{131}, pp.
  435--439, \doi{10.1103/PhysRev.131.435}.

\bibitem[{Riley \emph{et~al.}(2021)}]{Riley:2021pdl}
Riley, T.~E. \emph{et~al.} (2021). \enquote{{A NICER View of the Massive Pulsar
  PSR J0740+6620 Informed by Radio Timing and XMM-Newton Spectroscopy},}
  \emph{Astrophys. J. Lett.} \textbf{918}, 2, p. L27,
  \doi{10.3847/2041-8213/ac0a81},
  \href{http://arxiv.org/abs/2105.06980}{\UrlFont{arXiv:2105.06980
  [astro-ph.HE]}}.

\bibitem[{R\"opke \emph{et~al.}(1986)R\"opke, Blaschke and Schulz}]{Ropke:1986qs}
R\"opke, G., Blaschke, D.,  and Schulz, H. (1986). \enquote{{Pauli Quenching
  Effects in a Simple String Model of Quark / Nuclear Matter},} \emph{Phys.
  Rev. D} \textbf{34}, pp. 3499--3513, \doi{10.1103/PhysRevD.34.3499}.

\bibitem[{Shahrbaf \emph{et~al.}(2020{\natexlab{a}})Shahrbaf, Blaschke,
  Grunfeld and Moshfegh}]{Shahrbaf:2019vtf}
Shahrbaf, M., Blaschke, D., Grunfeld, A.~G.,  and Moshfegh, H.~R.
  (2020{\natexlab{a}}). \enquote{{First-order phase transition from
  hypernuclear matter to deconfined quark matter obeying new constraints from
  compact star observations},} \emph{Phys. Rev. C} \textbf{101}, 2, p. 025807,
  \doi{10.1103/PhysRevC.101.025807},
  \href{http://arxiv.org/abs/1908.04740}{\UrlFont{arXiv:1908.04740 [nucl-th]}}.

\bibitem[{Shahrbaf \emph{et~al.}(2020{\natexlab{b}})Shahrbaf, Blaschke and
  Khanmohamadi}]{Shahrbaf:2020uau}
Shahrbaf, M., Blaschke, D.,  and Khanmohamadi, S. (2020{\natexlab{b}}).
  \enquote{{Mixed phase transition from hypernuclear matter to deconfined quark
  matter fulfilling mass-radius constraints of neutron stars},} \emph{J. Phys.
  G} \textbf{47}, 11, p. 115201, \doi{10.1088/1361-6471/abaa9a},
  \href{http://arxiv.org/abs/2004.14377}{\UrlFont{arXiv:2004.14377 [nucl-th]}}.

\bibitem[{Shahrbaf \emph{et~al.}(2022)Shahrbaf, Blaschke, Typel, Farrar and
  Alvarez-Castillo}]{Shahrbaf:2022upc}
Shahrbaf, M., Blaschke, D., Typel, S., Farrar, G.~R.,  and Alvarez-Castillo,
  D.~E. (2022). \enquote{{Sexaquark dilemma in neutron stars and its solution
  by quark deconfinement},}
  \href{http://arxiv.org/abs/2202.00652}{\UrlFont{arXiv:2202.00652 [nucl-th]}}.

\bibitem[{Song \emph{et~al.}(2019)Song, Baym, Hatsuda and
  Kojo}]{PhysRevD.100.034018}
Song, Y., Baym, G., Hatsuda, T.,  and Kojo, T. (2019). \enquote{Effective
  repulsion in dense quark matter from nonperturbative gluon exchange,}
  \emph{Phys. Rev. D} \textbf{100}, p. 034018,
  \doi{10.1103/PhysRevD.100.034018},
  \url{https://link.aps.org/doi/10.1103/PhysRevD.100.034018}.

\bibitem[{Stone \emph{et~al.}(2021)Stone, Dexheimer, Guichon, Thomas and
  Typel}]{Stone:2019blq}
Stone, J.~R., Dexheimer, V., Guichon, P. A.~M., Thomas, A.~W.,  and Typel, S.
  (2021). \enquote{{Equation of state of hot dense hyperonic matter in the
  Quark\textendash{}Meson-Coupling (QMC-A) model},} \emph{Mon. Not. Roy.
  Astron. Soc.} \textbf{502}, 3, pp. 3476--3490, \doi{10.1093/mnras/staa4006}.

\bibitem[{Tanabashi \emph{et~al.}(2018)}]{ParticleDataGroup:2018ovx}
Tanabashi, M. \emph{et~al.} (2018). \enquote{{Review of Particle Physics},}
  \emph{Phys. Rev. D} \textbf{98}, 3, p. 030001,
  \doi{10.1103/PhysRevD.98.030001}.

\bibitem[{Terazawa(1979)}]{Terazawa:1979}
Terazawa, H. (1979). \enquote{Prediction of a super-hypernucleus, $h_{\lambda}$
  (hexalambda), with $b=6$ and $s=-6$ at $m\cong 5.6\sim 6.3gev$ in the mit bag
  model and at $m\cong 7.0gev$ in the quark-shell model,} Tech. Rep. INS-336,
  Tokyo University, Tokyo, Japan.

\bibitem[{Typel(2005)}]{Typel:2005ba}
Typel, S. (2005). \enquote{{Relativistic model for nuclear matter and atomic
  nuclei with momentum-dependent self-energies},} \emph{Phys. Rev. C}
  \textbf{71}, p. 064301, \doi{10.1103/PhysRevC.71.064301}.

\bibitem[{Typel(2018)}]{Typel:2018cap}
Typel, S. (2018). \enquote{{Relativistic Mean-Field Models with Different
  Parametrizations of Density Dependent Couplings},} \emph{Particles}
  \textbf{1}, 1, pp. 3--22, \doi{10.3390/particles1010002}.

\bibitem[{Typel \emph{et~al.}(2010)Typel, R{\"o}pke, Kl{\"a}hn, Blaschke and
  Wolter}]{Typel:2009sy}
Typel, S., R{\"o}pke, G., Kl{\"a}hn, T., Blaschke, D.,  and Wolter, H.~H.
  (2010). \enquote{{Composition and thermodynamics of nuclear matter with light
  clusters},} \emph{Phys. Rev. C} \textbf{81}, p. 015803,
  \doi{10.1103/PhysRevC.81.015803},
  \href{http://arxiv.org/abs/0908.2344}{\UrlFont{arXiv:0908.2344 [nucl-th]}}.

\bibitem[{Typel and Wolter(1999)}]{Typel:1999yq}
Typel, S. and Wolter, H.~H. (1999). \enquote{{Relativistic mean field
  calculations with density dependent meson nucleon coupling},} \emph{Nucl.
  Phys. A} \textbf{656}, pp. 331--364, \doi{10.1016/S0375-9474(99)00310-3}.

\bibitem[{Witten(1984)}]{Witten:1984rs}
Witten, E. (1984). \enquote{{Cosmic Separation of Phases},} \emph{Phys. Rev. D}
  \textbf{30}, pp. 272--285, \doi{10.1103/PhysRevD.30.272}.

\end{thebibliography}


\end{document}